\documentclass[letterpaper]{article}
\usepackage{aaai24}
\usepackage{times}  
\usepackage{helvet}  
\usepackage{courier}  
\usepackage[hyphens]{url}  
\usepackage{graphicx} 
\usepackage{color} 
\def\UrlFont{\rm}  
\usepackage{natbib}  
\usepackage{caption} 
\usepackage{subcaption}
\usepackage{hhline}
\usepackage{booktabs}
\usepackage{multirow}
\usepackage{amsmath}
\usepackage{url}
\usepackage{float}

\usepackage{amssymb}
\usepackage{tabularx}
\usepackage{makecell}
\usepackage{multicol}
\usepackage{tipa}

\usepackage[T1]{fontenc}

\usepackage[utf8]{inputenc}

\usepackage{microtype}

\usepackage{inconsolata}

\newcommand{\figref}[1]{Figure \ref{#1}}

\newcommand{\tabref}[1]{Table \ref{#1}}

\usepackage[ruled,linesnumbered]{algorithm2e}
\usepackage{newfloat}
\usepackage{listings}
\DeclareCaptionStyle{ruled}{labelfont=normalfont,labelsep=colon,strut=off} 
\floatstyle{ruled}
\newfloat{listing}{tb}{lst}{}
\floatname{listing}{Listing}
\title{Towards Lexical Analysis of Dog Vocalizations via Online Videos}
\author {
    Yufei Wang\textsuperscript{\rm 1},
    Chunhao Zhang\textsuperscript{\rm 2},
    Jieyi Huang\textsuperscript{\rm 3},
    Mengyue Wu\textsuperscript{\rm 4},
    Kenny Zhu\textsuperscript{\rm 5*}
}
\affiliations {
    \textsuperscript{\rm 1,2,3,4}Shanghai Jiao Tong Univerisity\\
    \textsuperscript{\rm 5}The University of Texas at Arlington\\
    \textsuperscript{\rm 1}arthur-w@sjtu.edu.cn, \textsuperscript{\rm 2}forest\_zch@sjtu.edu.cn, \textsuperscript{\rm 3}xsiling1@gmail.com\\
    \textsuperscript{\rm 4}wumengyue@cs.sjtu.edu.cn, \textsuperscript{\rm 5}kenny.zhu@uta.edu
}

\begin{document}

\maketitle
\begin{abstract}
Deciphering the semantics of animal language has been a grand challenge.
This study presents a data-driven investigation into the semantics of dog vocalizations via correlating different sound types with consistent semantics. 
We first present a new dataset of Shiba Inu sounds, along with contextual information such as location and activity, collected from YouTube with a well-constructed pipeline. The framework is also applicable to other animal species. Based on the analysis of conditioned probability between dog vocalizations and corresponding location and activity, we discover supporting evidence for previous heuristic research on the semantic meaning of various dog sounds. For instance, growls can signify interactions. Furthermore, our study yields new insights that existing word types can be subdivided into finer-grained subtypes and minimal semantic unit for Shiba Inu is word-related. For example, whimper can be subdivided into two types, attention-seeking and discomfort. 
 
\end{abstract}

\section{Introduction}
\label{sec:intro}
Animal languages have captured the curiosity of scientists for years and animals use vocal expressions to communicate ~\cite{garcia2017animal}.
Despite various attempts from diverse perspectives, deciphering the intricately complex semantic meanings within animal communication systems remains a challenge ~\cite{andreas2022towards,scott2023animal}. Acquiring a deeper comprehension of animal language holds significant implications for unraveling their social structures, and intelligence, and facilitating human-animal interactions. 

Within the expansive realm of animal languages, the study of \textbf{dog language} holds particular interest. Dogs, as one of the most popular and widely kept pets, engage in constant interaction with human beings through their vocal expressions.
Given the massive interactions between dogs and people during the domestication process, dogs' vocal behavior undergoes considerable changes ~\cite{jieyiacl2023, feddersen2000vocalization} and it is reasonable to infer that diverse sounds emitted by dogs in varying scenes carry distinct significances. Previous works on dog language have largely relied upon experimental knowledge and heuristic
subjective observations, which depend on long-term experiences and costly data-collection and can be limited and prone to biases ~\cite{yin2002new,pongracz2010barking,farago2017dog}. Only coarse-grained meanings can be drawn given limited data and corresponding scenes from these studies. Our web-data-driven exploration leverages a more comprehensive methodology to uncover finer-grained semantics with a broader context. The utilization of \textbf{web data} offers a wealth of information, introducing numerous possible variables and semantic clues, thus enabling a more comprehensive analysis. 

\begin{figure*}[t]
	\centering
	\includegraphics[width=0.85\textwidth]{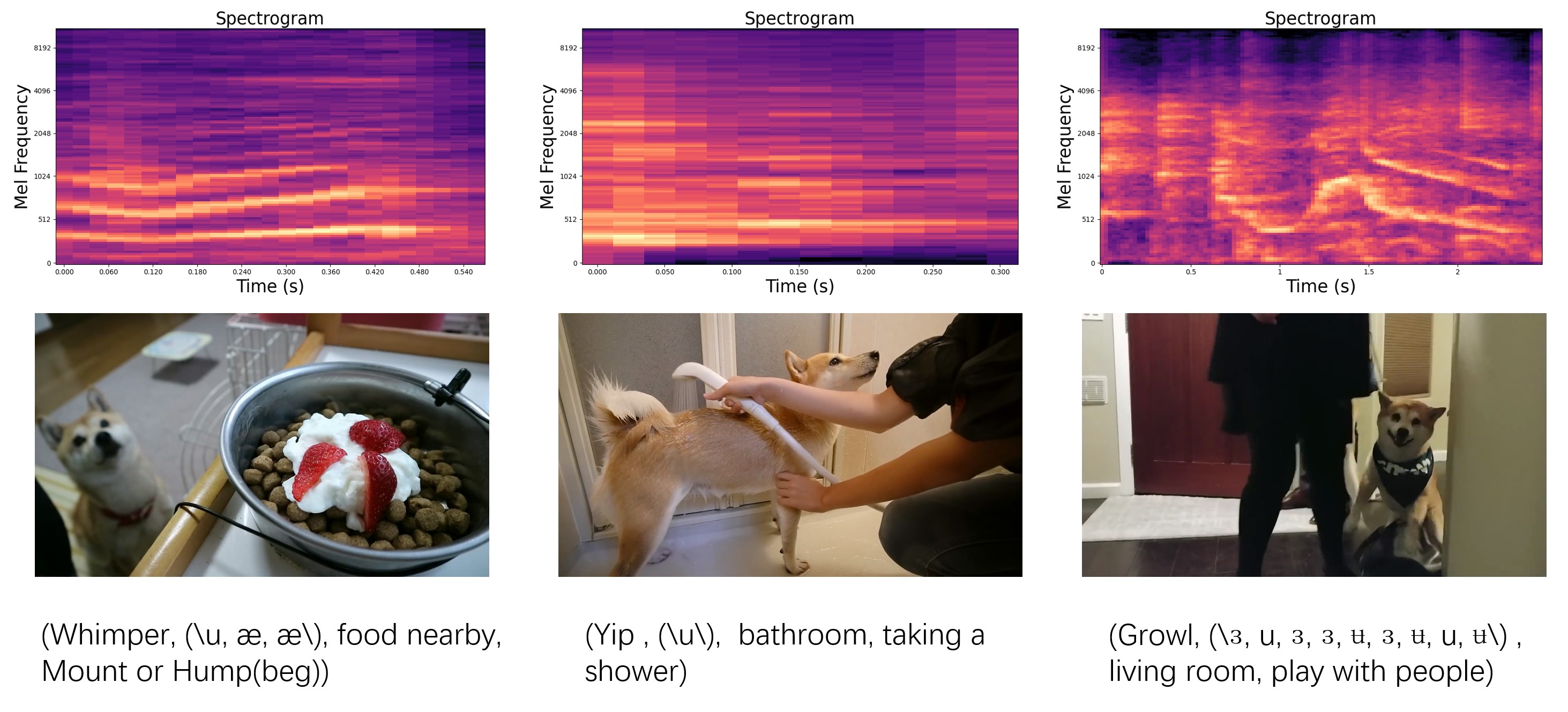}
	\caption{Introducing scenario: Spectrograms of dog sounds in the corresponding context. When the dog is begging for food, taking a shower, or playing with people, its vocal spectrograms differ a lot. We map possible words and subwords extracted from dog barking audio with location and activity, forming our quadruplet dataset.  
\texttt{<word, IPA symbols, location, activity>}.
	} 
	\label{fig:intropic}
\end{figure*}

There is a wide range of vocalizations dogs can produce ~\cite{yeon2007vocal}, which are affected by various engaged objects, the emotion of the dog, the surrounding environment the dog is located in, the activity the dog is doing, the object the dog interacts with, and even the age and gender of the dog may play a key factor ~\cite{pongracz2005human,molnar2009dogs}. 
Given the benefits of using web data, we opt to focus on the Shiba Inu breed for our research as Shiba Inu is a widely adopted dog at home and plenty of video data is available on YouTube. Hereby, we investigate the semantic meaning of the Shiba Inu dog sound according to two important factors of the context, the \textbf{location} and the \textbf{activity} of the dog as these two factors are currently available. 

 

To understand the semantic meaning of dog language, previous works always record dog sounds in different scenarios and then analyze them. As we have online videos, there are several challenges: extract the meaningful dog sound and context from videos and ascertain if they show consistent patterns with context. In this work, we implement the first data-driven, evidence-based research sourced from social media to give fine-grained semantics to understand the vocal language of Shiba Inu. We construct data as ~\figref{fig:intropic} for our dataset to map dog sound with context which enables us to take a closer look into the behaviors of animals and some hidden \textbf{semantic patterns}. For example, one kind of dog sound, bow-wow, is usually mixed with bark by previous researchers. However, in our work, we find that it shows the curiosity of dogs for the surroundings while bark does not exhibit such meaning. 
On the other hand, our method has more detailed contexts for analyzing animal languages, which will benefit future works. We identified as many as 11 locations and 14 activities, in which dogs might produce 6 different vocal sounds to signify various meanings. To associate vocal expression patterns with possible lexical meanings, we need to respectively extract vocal sounds and their transcriptions as well as the activity and location that might give rise to a change of meanings. With these fine-grained labels, we can capture subtle semantic differences and constant vocal patterns under these contexts to explore the semantics of vocalizations for Shiba Inu dogs.
Our main contributions can be summarized as follows:

\begin{itemize}
	\item We propose a universal pipeline to process and analyze dog-related videos on YouTube to understand the lexical semantics of dog language. The framework is reusable to other animal species for which videos are available.
\item We are the first to implement data-driven research to study dog semantic language from web data. We build a dataset of 10,779 quadruplets that contains 6 distinct words, subwords, and corresponding context for exploring dog language. We define fine-grained 14 activities and 11 locations that could imply different semantic meanings which can be extracted from videos. 
\item Through our investigation of dog sound patterns, we have uncovered several conclusions that align with existing human knowledge and previous research. Additionally, we have gained some unique insights that have been under-explored. 
\end{itemize}


\section{Problem}
\label{sec:problem}

Our goal is to understand the lexical semantics of dogs and explore the minimal semantic unit.  
We seek to address the following technical problems: 
\begin{enumerate}
	\item Do dogs use consistent vocal patterns to signify certain meanings?
	\item How to compute the correlation between vocal expressions with possible factors that give rise to different certain meanings?


\end{enumerate}

To answer these questions, we need to classify distinct sound types, which are defined as ``words'' and we further phonetically transcribe these words, which are signified as subwords in Section \ref{sec:divide}. Regarding contextual information to uncover the semantics, we define a diversed and comprehensive list for location and activity and utilize respective extraction methods in Section \ref{sec:infer_context}.

\section{Approach}
\label{sec:approach}

\begin{figure}[th]
	\centering
	\includegraphics[width=0.9\columnwidth]{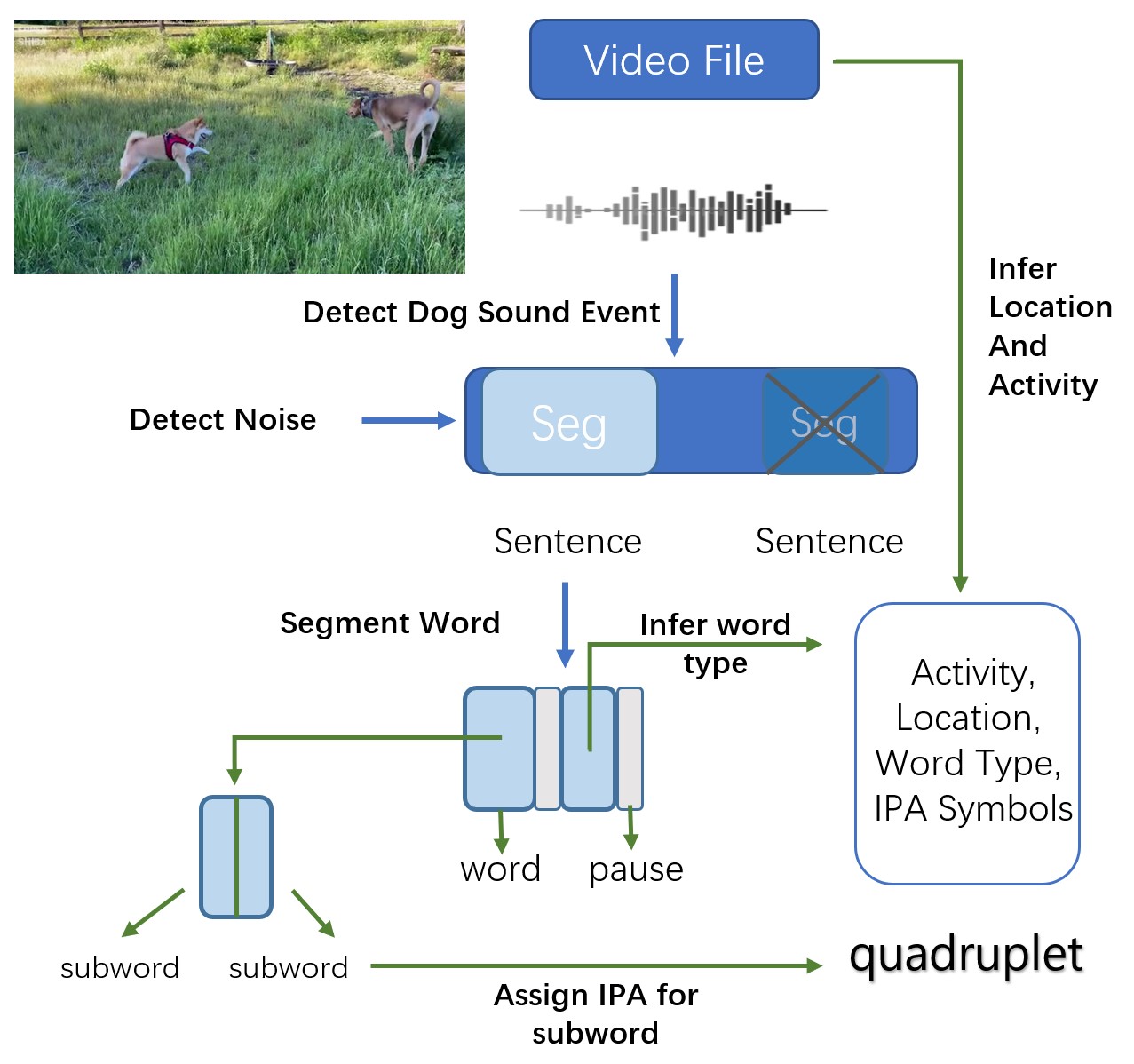}
	\caption{The overall view of pipeline.}

\label{fig:method}
\end{figure}

In this section, we present our pipeline (as shown in ~\figref{fig:method}) including data collection, vocalization processing procedures (word segmentation, subword extraction, and phonetic transcription), as well as contextual information extraction methods (location and activity recognition). Implementation details can be found in Appendix A.

\subsection{Collect Shiba Inu Online Videos}
\label{sec:collect_data}
We first use the keyword ``Shiba Inu'' to identify users who own such dogs and only upload shiba inu videos with dog sounds then we collect up to 13,164 Shiba Inu related videos posted by these users within days which can be easily expanded in the future.

\subsection{Sentence Split, Word Segmentation and Subword Extraction}
\label{sec:divide}
Once videos are downloaded, we begin to process the audio tracks.  Similar to human language, where words serve as the fundamental units for constructing sentences, we hypothesize that a similar concept can be applied to dog language. We define a \textbf{``word''} as an independent and contiguous dog vocal sound that typically lasts around 1 second, and it is bounded by some noticeable pauses. A \textbf{``sentence''} is a sequence of consecutive words. A \textbf{``subword''} is a subpart of ``word'' and is represented by an IPA symbol.


\paragraph{Sentence Extraction}

\begin{algorithm}[]
  \SetAlgoLined
  \KwData{Audio tracks}
  \KwResult{``sentences'' of dogs }
  \While{pass accross segments of audio track}
  {
    feed the audio segment feature into PANNs\;
    \eIf{PANNs infers as dog sound without accompanying noise}
    {
      	audio segment belongs to a ``sentence''\;
    }
	{
      pass to next segment\;
  }
  }
  \caption{Sentence Extraction}
\end{algorithm}

To extract the word clips, we adopt a similar methodology as \citet{jieyiacl2023}. The initial step is detecting dog vocalizations and splitting the audio tracks into ``sentences'', continuous sequences of dog vocal sounds. We pass through the frames of audio and determine whether they contain dog sounds. If so, we extract these audio clips and then we remove noise other than dog vocalizations to ensure that there is no accompanying music or human speech in the background. We apply PANNs~\cite{kong2020panns}, a sound event detection model that is pre-trained on the large-scale Audioset~\cite{gemmeke2017audio} dataset with 527 sound classes. 


\paragraph{Word Segmentation}
\begin{algorithm}[]
  \SetAlgoLined
  \KwData{``Sentences''}
  \KwResult{``Words''}

  \While{pass through segments of ``sentences''}
  {
    Feed segment feature into PANNs\;
    \eIf{PANNs infers as ``dog'' other than ``silence'' }
    {
      segment belongs to a ``word''\;
    }
	{
      Use segment to split ``words''\;
  }
  }
  \caption{Word Segmentation}
\end{algorithm}

These ``sentences'' may contain short pauses in the middle that can be used to separate the words and the next step is to segment ``words'' in ``sentences''. We finetune PANNs to determine each start and end of a single ``word'' by detecting a frame in the audio clip which transits from a silence frame to a dog frame with a gap of 0.1 seconds between frames. We manually create labels for the event ``dog'' with a total data length of 715 seconds. This finetuned model is capable of detecting the small pauses within ``sentences'' and extracting the singular ``words''. Lastly, we follow the definition of 6 different dog vocalization patterns defined in Audioset, which are distinct from each other. We labeled 240 audio clips and trained a classification model.  

\begin{equation}
\text{Wordtype} = \arg\max_{i} P_{PANNs}(\text{soundtype}_i | \text{word}).
\end{equation}


\paragraph{Subword Segmentation and Phonetic Transcription}
We first split ``words'' into ``subwords'' with the method as ~\cite{rasanen2018pre}, which uses sonority fluctuation in audio to segment words into smaller units. Thus, a word can be further split into smaller parts according to the sonority changes. In order to present the vocal characteristics of each subpart, we phonetically transcribe each subword with international phonetic alphabet (IPA) symbols. We compute the acoustic feature distances with standard pronunciations ~\cite{ipa0, ipa1}. In this way, each subword is represented by an IPA vowel symbol, and a ``word'' is represented by a sequence of IPA symbols. 

\subsection{Surrounding Context Extraction}
\label{sec:infer_context}
Since images and sound are naturally aligned in a video,  we can extract the \textit{location} and \textit{activity} of the dog while it utters a particular ``word'', from the image frames that synchronize with the audio frames. The end product of this phase will be a sequence of quadruplets  consisting of \texttt{<word, subwords, location, activity>}, extracted from the videos. 

 \paragraph{Location} 
Given the begin and end timestamp of a ``word'', five image frames in that time range are sampled, then sent into an image classification model to determine the location of the dog when the ``word'' occurs. Here we finetune the pre-trained model from~\citet{zhou2017places} trained on the scene-centric datasets Places365 based on thousands of pictures sampled from Shiba Inu videos to predict the dog's location. The fine-tuned model achieves 77.99\% on Top-1 accuracy and 97.13\% on Top-5 accuracy. 
 
We denote the following variables:
$t_{\text{word}}$: Timestamp of the ``word'' occurrence between begin and end, 
$I_{\text{i}}$: An image from a batch of images constructed from five frames sampled around the timestamp $t_{\text{word}}$,
$M_{\text{class}}$: The class predicted by the model given $I_{\text{i}}$,
$L_{\text{dog}}$: Location of the dog when the ``word'' occurs,
$\text{Majority Vote}$ is a function that selects the class with the highest frequency among the predicted class of the individual images in the batch.
The formula for location inference is as:
\begin{equation}
L_{\text{dog}} = \text{Majority Vote}\left(\bigcup_{i=1}^{n} M_{\text{class}}(I_i)\right)
\end{equation}

\paragraph{Activity} To get the activity information about the dog, we sample a five-second video clip based on the timestamp of the ``word'' that we chooce a range of (begin timestamp  - 1s, end timestamp + 1s), and then a video understanding model is applied to decide what the dog is doing. The pre-trained model Temporal Segment Networks~(TSN)~\cite{wang2018temporal} performs well on the task of video understanding. In this study, we annotated 2534 video clips sampled from the Shiba Inu videos and finetune the pre-trained model. The fine-tuned model achieves 61.40\% on Top-1 accuracy and 92.40\% on Top-5 accuracy. Based on these two fine-tuned models, given the timestamp of a ``word'', we can explore the location and activity from the video.


\section{Dataset}
We obtain a large-scale timestamp-aligned dataset of \texttt{<word, subwords, location, activity>}. We present the details of our dataset and quality of it in the following part. 

\subsection{Statistics of the dataset}
\label{sec:dataset}
The dataset contains 10,779 quadruplets from corresponding 3,068 videos and 5,834 sentences, including 6 different types of dog sounds, 11 location categories, 14 activity categories, and 20 IPA vowels for subwords as shown in ~\tabref{tab:dataset}. We follow the definitions of dog vocalization types defined in Audioset and we adopt the most frequent locations and activities tailored for Shiba Inu by combining commonsense and data manual checking for location and activity. More details of the dataset can be seen in Appendix C. 


\begin{table}[th]
	\small
	\begin{tabular}{p{0.15\columnwidth}|p{0.7\columnwidth}}
		\toprule
		\textbf{Context} & \textbf{Possible values} \\ \midrule
		Word Type & Bow-wow, Bark, Whimper, Growl, Howl, Yip \\ \midrule
		Phonetic Symbol & \textipa{\textbackslash u,æ,\textrevepsilon,\textbaru,\textturnm,\textturna,a,\textbaro,\textscoelig,\textschwa,\textupsilon,\textsci,\textturnscripta ,\textepsilon,\textturnv,\textscripta,œ,\textopeno,\textbari,e\textbackslash} \\ \midrule
		Location & Living Room, Food Nearby, Grass, Cage, Road, Bathroom, Snowfield, Beach, Square, Vehical Cabin, Others    \\ \midrule
		Activity & Mount Or Hump (beg), Play With People, Sit, Lay Down, Walk, Sniff, Eat, Stand, Take a Shower, NoDog, Run, Be Touched,  Unknown, Fight With Dogs, Show Teeth or Bit\\ 
		\bottomrule
	\end{tabular}
	\caption{Respective categories for dog vocalizations and contextual information in our dataset.} 
	\label{tab:dataset}
\end{table}

\begin{figure*}[h]
\centering
\includegraphics[width=1.8\columnwidth]{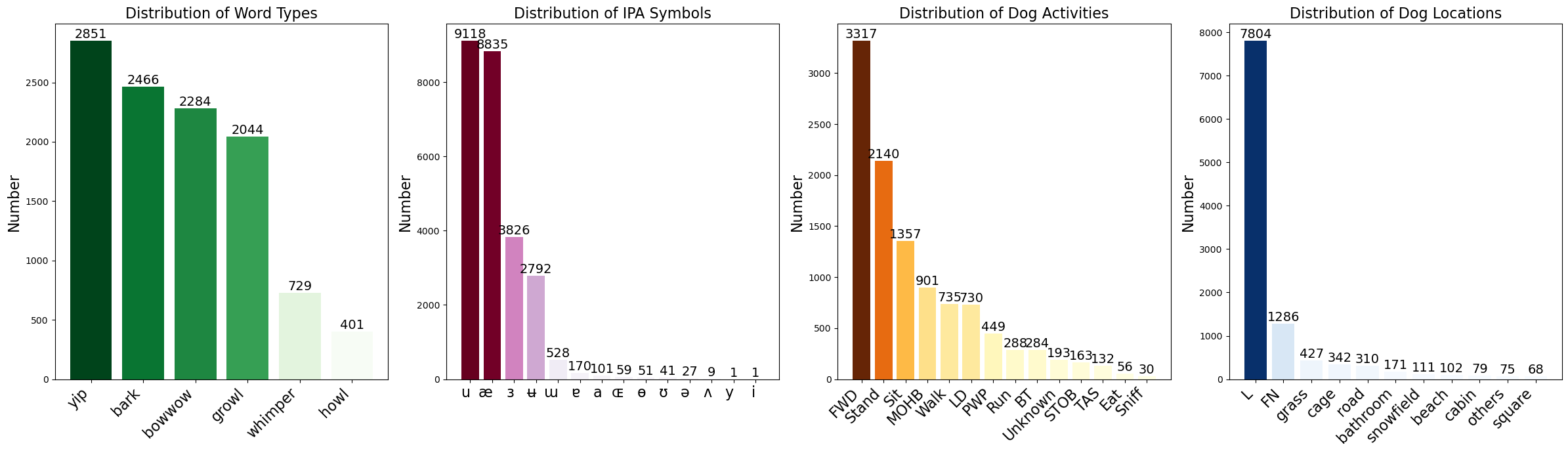}
\caption{Prior distributions of word types(green), IPA symbols(pink), locations(blue) and activities(red) in
the quadruplet sequences. The numbers shown are the number of times for different word types, locations, and activities.
``L'', ``FN'', ``FWD'', ``MOHB'', ``LD'', ``PWP'', ``BT'', ``STOB'', ``TAS'' represent  ``Livingroom'', ``Food Nearby'', ``Fight With Dogs'', ``Mount Or Hump (Beg)'', ``LayDown'', ``Play With People'', Be Touched'', ``Show Teeth Or Bit'', ``Take A Shower'' respectively. 
}
\label{fig:distri}
\end{figure*}

~\figref{fig:distri} shows the prior distribution of the items in the dataset across different word types, subword IPA symbols, location types, and activity types. This data imbalance reflects real-world data distribution, that pet dogs kept at home are more prone to stay in the living room and exhibit activities like standing and fighting with dogs. We consider this imbalanced prior distribution when computing the correlations between them. 


\subsection{Quality of the dataset}
We present the accuracy of each step to ensure the high quality of our dataset. For word segmentation and word classification, we randomly sample 200 segmented words to listen to. Three annotators have to label whether this is a singular dog word and whether the word is correctly classified. Their respective accuracy is 0.95 and 0.84. We achieve an accuracy of 0.78 for location classification on our manually labeled pictures and an accuracy of 0.59 for the top 1 and 0.9416 for the top 5 for activity classification.

\section{Results and Analysis}
\label{sec:results}
To answer the first question, we analyze the relationship between words and context. We then explore the relationship between words and subwords. We present a summarization of findings that uncovers dog vocalization patterns. To answer the second question, we adopt conditioned probability. 
 
\subsection{Correlation between word type and location}
\begin{figure*}[h]
\centering
\begin{subfigure}[]{0.4\textwidth}
	\centering
	\includegraphics[width=0.86\textwidth]{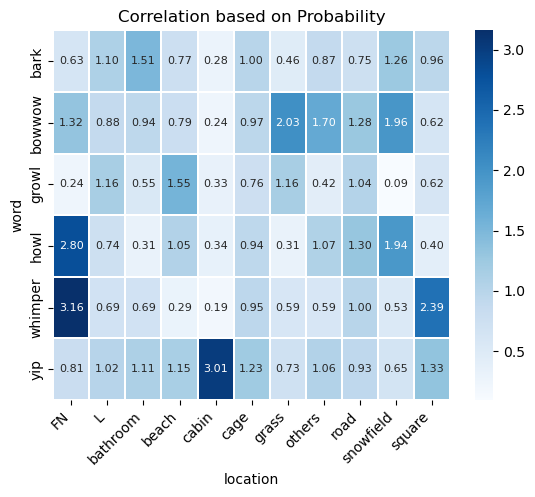}
	\caption{Correlation between word types and locations.}
	\label{fig:sound_location}
\end{subfigure}
\begin{subfigure}[]{0.4\textwidth}
	\centering
	\includegraphics[width=0.86\textwidth]{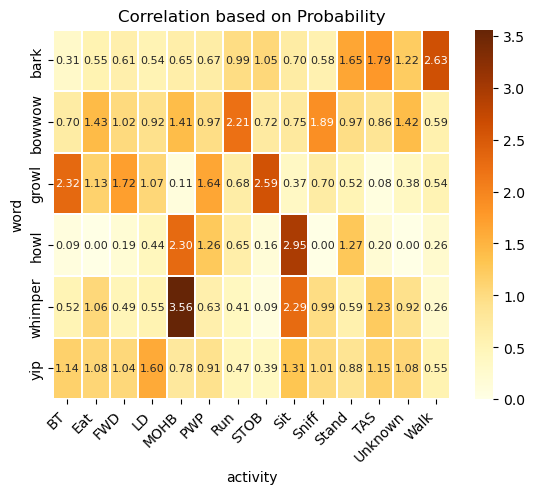}
	\caption{Correlation between word types and activities.}
	\label{fig:sound_activity}
\end{subfigure}

\begin{subfigure}[]{0.4\textwidth}
	\centering
	\includegraphics[width=0.86\textwidth]{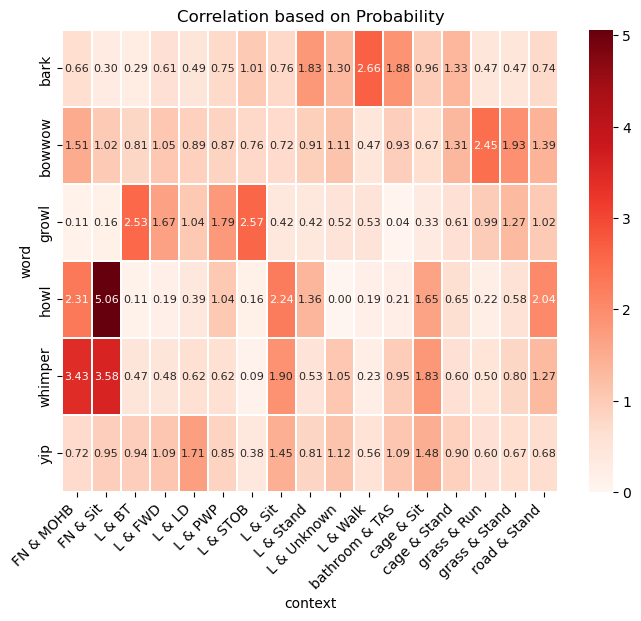}
	\caption{Correlation between word types and contexts.}
\label{fig:one_sound_context}
\end{subfigure}
\begin{subfigure}[]{0.4\textwidth}
	\centering
	\includegraphics[width=0.86\textwidth]{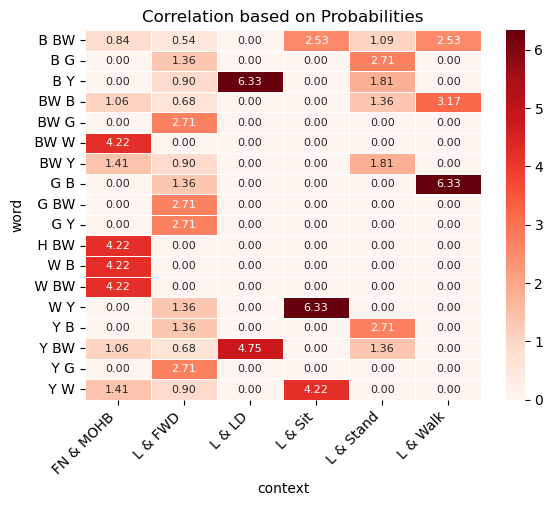}
	\caption{Correlation between bi-gram of words and contexts. 
``B'', ``BW'', ``G'', ``H'', ``Y'', ``W'' represent ``bark'', ``bow-wow'', ``growl'', ``howl'', ``yip'' and ``whimper'' respectively.}
	\label{fig:sequenc_sound_context}
\end{subfigure}
\caption{Correlation to explore semantics of words.}
\label{fig:correlation}
\end{figure*}

The vocalization of a dog varies depending on its location. Based on the dataset, we compute the correlation between location and word as shown in ~\figref{fig:sound_location}: 
\begin{equation}
\frac{P (\text{location}| \text{word})}{P(\text{location})}
\end{equation}
By dividing each item in the formula by its prior probability, we can offset the influence of the original umbalanced frequencies of the word types and locations. 



\paragraph{``Bow-wow'' can be used to express curiosity.}
Bow-wows are usually used when they are outside and in unfamiliar surroundings like snowfields, grass, and other places. This word may implicate an exploration of the environment. 
\paragraph{``Whimper'' can be seen as attention-seeking when food nearby.}
``Whimper'' can be used to elicit people attention ~\cite{handelman2012canine}. When there is food nearby, they may beg for food.

\subsection{Correlation between word type and activity}
We use the same method to analyze the relationship between word types and 
activities in \figref{fig:sound_activity} as: 
\begin{equation}
\frac{P (\text{activity}| \text{word})}{P(\text{activity})}
\end{equation}
\paragraph{``Bow-wow'' may indicate movements or food.} Dogs exhibit a preference for using the ``bow-wow'' sound when engaging in activities that involve movement, such as walking and sniffing. ``Bow-wow'' may also be a signal for food as begging for food or eating. 
\paragraph{``Growl'' expresses interation with outside.}
``Growl'' appears to be used to express interactions ~\cite{handelman2012canine}. It is prevalent when be touched, fight with dogs, play with people and show teeth or bite. 
\paragraph{``Whimper'' and ``howl'' are used when relatively steady.}
Dogs are usually howling when sitting down and standing. We find that Shiba Inu tends to ``howl'' when they are begging for food which is a new observation. 
Dogs often ``whimper'' when engaged in activities like sitting and begging for food. It may indicate contact seeking and a kind of submission ~\cite{pongracz2010barking}.


\subsection{Correlation between word type and context}
When putting location and activity together as in ~\figref{fig:one_sound_context}:
\begin{equation}
\frac{P (\text{(location, activity)}| \text{word})}{P(\text{(location, activity)})}
\end{equation}
Because of the numerous combination possibilities for context, we define a threshold value of 100, indicating that a particular combination is considered statistically significant only if it occurs more than 100 times. ~\figref{fig:one_sound_context} depicts this relationship between word type and context.
\paragraph{``Howl'' and ``whimper'' come up frequently with food.} ``Whimper''represent attention-seeking, thus it can be interpreted as attention for food when ``food nearby'' and the dog ``begs''. ``Howl'' is used when sitting and food nearby. 
\paragraph{``Yip'' and ``whimper'' can express discomfort.}  They can express pain or discomfort ~\cite{yeon2007vocal,web2018yip}. When the dog is located in the cage and laying down, it tends to yip and whimper.  
\paragraph{``Bark'' can express discomfort.} Shiba Inu barks when it is taking a shower in the bathroom. This can express its discomfort and warning ~\cite{yeon2007vocal} because dogs usually don't like baths. 


\subsection{Analysis of the sequence of quadruplets} 
To analyze quadruplet sequences, we start by picking words in the same sentence where the word type changes. We hypothesize that a sequence of different word types in a similar context carries a specific meaning. We only consider combinations that appear frequently, setting a threshold of 10 to distinguish patterns from chance occurrences. We explore the formula as shown in ~\figref{fig:sequenc_sound_context}:
\begin{equation}
\frac{P (\text{(location, activity)}| (w_1;w_2))}{P(\text{(location, activity)})}
\end{equation}
while $(w_1;w_2)$ represents two consecutive words with different types, which is defined as a \textit{bi-gram} here. We highlight several insights. Our findings suggest that the sequence of words might not strongly indicate distinct semantic meanings. For instance, comparing "Bark Bow-wow" and "Bow-wow Bark" reveals differing context distributions. However, in the cases of "Whimper Bow-wow" and "Bow-wow Whimper", we observe similar distributions despite the word order variation.

\begin{figure}[t]
\centering
\begin{subfigure}[]{0.4\textwidth}
	\centering
	\includegraphics[width=0.85\textwidth]{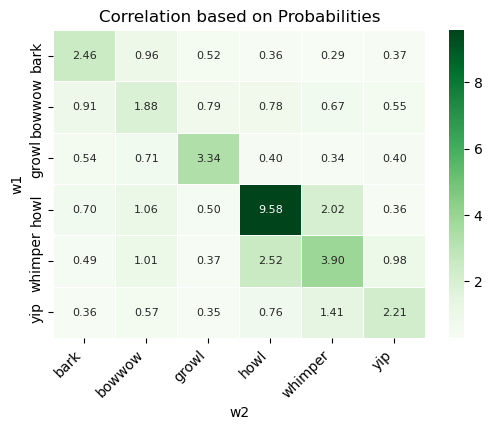}
	\caption{Correlation based on probability for P($w_2$|$w_1$).}
	\label{fig:w2w1}
\end{subfigure}
\begin{subfigure}[]{0.4\textwidth}
	\centering
	\includegraphics[width=0.9\textwidth]{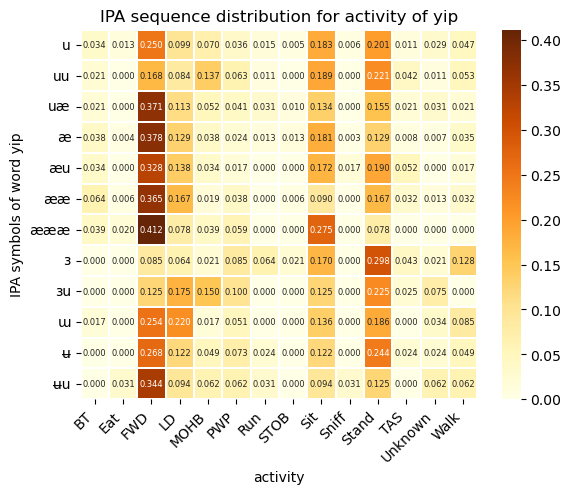}
	\caption{Activity distribution of most frequent subword combinations for yip.}
	\label{fig:subwords_yip}
\end{subfigure}
\caption{Exploring sequence of words and activity distribution of subwords for yip}
\end{figure}

To provide additional evidence, we examine the bi-gram probability of two-word sequences as $w_1$ followed by $w_2$, where $w_1$ and $w_2$ represent two words. To enhance clarity, we normalize each item by dividing it by the prior distribution of $w_2$, revealing relative changes. The result for: 
\begin{equation}
{P (w_1|w_2)}
\end{equation}
is shown in ~\figref{fig:w2w1}. As observed, ``howl'' and ``whimper'' are highly probable of following each other and this implies their semantics are kind of overlapping. Same for ``bark'' and ``bow-wow''. This shows the words contain multiple semantics.



\subsection{Analysis of subwords semantics}
Previous works show for one word type, it conveys multiple meanings. Each word contains one or multiple different subwords. By combining analyzing the multiple meanings for one word type and its possible subwords, we prove that one word type can be further divided into finer-grained types, which is the first data-driven experimental proof. For subwords in the same word, we first collect the most frequent subwords that can be one or a sequence of IPA vowel symbols, then we illustrate their distribution on the context and they vary a lot, which means that they convey different semantics. As shown in ~\figref{fig:subwords_yip}, we present the activity distribution of the most frequent subwords for yip. As we can see, for the same scene fight with dogs, different IPA symbol sequence shows an dissimilar distribution. 

We further observe that these frequently different subwords are not restricted to one dog, which means this semantic difference is not by accident or caused by specific characters of one dog. A summary table is appended in Appendix C. By combining analysis of words and subwords, we explore that the distribution of the IPA symbol is mainly influenced by the distribution of the word which contains this symbol and this illustrates that the minimal semantic unit is word-related instead of subword. A detailed explanation is in Appendix D.


\begin{table}[t]
\small
\centering
\begin{tabular}{p{0.13\columnwidth}|p{0.45\columnwidth}|p{0.3\columnwidth}}
\toprule
\textbf{Lexical Symbols} & \textbf{Previous Meaning} & \textbf{Our Meaning}\\
\hline
Whimper & attention-seeking \cite{handelman2012canine},
				discomfort \cite{web2018dog} & attention-seeking, beg for food, steady, discomfort \\
\hline
Yip & discomfort ~\cite{web2018yip} & loneliness \\
\hline
Bow-wow & NA & show curiosity, movement\\
\hline
Growl & playing interaction ~\cite{handelman2012canine} & interation with outside \\
\hline
Bark & warning \cite{handelman2012canine} & discomfort \\
\hline
Howl & warning, play, group cohesion ~\cite{ani8080131} & signal for food, steady \\
\bottomrule
\end{tabular}
\caption{Summary of findings with comparison to previously-discovered meanings.}
\label{tab:finding}
\end{table}

\subsection{Comparison with Previous Works}
We present our findings including the evidence to support previous theoretical studies and our new observations in \tabref{tab:finding}.
Through the analysis, we found most of the patterns we mentioned are consistent with the previous qualitative research. We also discover more detailed interpretations of those existing patterns and provide possible new findings. We are the first web-data-driven approach to analyze the minimal semantic unit of the Shiba Inu dog language.

\section{Related Work}
\label{sec:related}

It has long been difficult to understand what animals are trying to express. Since we can't speak to them directly, just trying to understand their language becomes an explorable target. Early studies have contributed to our understanding of dogs: dog vocalizations in different context~\cite{molnar2009dogs, robbins2000vocal}, emotion recognition through vocal cues~\cite{pongracz2006acoustic}, and the development of image and video analysis techniques for pet understanding~\cite{mao2023pet}. These studies give us a basic understanding of dog sounds. However, they either only conduct testing experiments, or only studied images and sound signals without mining the relationship between them. Our study does lexical analysis and connects it with the goal of understanding dogs.


In our communication with dogs, visual signals complement our understanding through sound signals. There are some interesting datasets that encourage visual tasks which help us understand dogs, such as first-person videos from a camera on the back of dogs for activity classification~\cite{iwashita2014first}, videos with skeletons labeled which help to detect poses~\cite{cao2019cross} and a collection of videos about different animal behavious~\cite{ng2022animal}. These studies have greatly broadened the methods for animal action recognition, but there is a lack of a dataset for dog activities in domestic scenes.


In addition to dogs, some animals are social animals and have a lot of interspecies interactions and it is an interesting topic about how they communicate with each other. Cetaceans~\cite{bermant2019deep}, elephants~\cite{rossman2020contagious} have a high degree of social complexity, and acoustic features can be used for detection or analyzing responses to other signals. Voice of birds~\cite{koh2019bird, salamon2017fusing, adavanne2017stacked} also brings information that can be used to detect birds or classify breeds. However, it is expensive to record sound and the restricted experiment environment restricts the diversity of data. Our data-driven research pipeline takes advantage of the vast amount of data available on the Internet to build a scalable and diverse dataset.


Also, it is of vital importance to make sure the boundary of the words are precise and the background is clear. AudioSet~\cite{gemmeke2017audio} consists of hundreds of audio event classes with human-labeled sound clips and further study~\cite{hershey2021benefit} collected frame-wise labels for a portion of the AudioSet to improve the detection performance. Based on that large-scale audio event dataset, PANNs~\cite{kong2020panns}, including several models for sound event detection are pre-trained. Previous research also tries to decipher different dog sounds ~\cite{web2023sounds, web2018yip}. Our work proposes the definition of words and develops word segment methods based on these foundation works. 

By further splitting the words, we also explore the semantics of subwords and the minimal semantic unit for Shiba Inu dog language. The results show that subwords expressed by IPA vowels do not show a special meaning. This could be attributed to the possibility that IPA vowels are tailored for human language rather than animal communication, yielding the potential need for a broader range of symbols to accurately capture the nuances in vocalization differences.


\section{Conclusion}
In this paper, we introduce a data-driven approach for exploring the semantics of Shiba Inu vocalization and constructing a dataset including dog words and corresponding context. Compared to the former approaches, it is cost-saving and extensible for datasets. Due to the large amount of data, it provides a probability to explore new contexts and find fine-grained semantics. The approach can be transferred to other animals easily. 

We also make some preliminary observations and analyses on the dataset. The analysis shows that the different dog words are used in various contexts. Most of our findings are consistent with previous research and we also explore new semantics for words. We present the word type conveys multiple meanings and can be further divided into more fine-grained types. By given evidence, we declare the minimal semantic unit for the Shiba Inu dog language is word-related. For future work, we can further classify dog sounds into more fine-grained types because we realize a word conveys multiple meanings.

\bibliography{anthology,custom}


\end{document}


\maketitle
\appendix

\section{Implementation details}
In this part, we explain our implementation details. For PANNs, we use the original model and predefined parameters to extract ``sentences''. Based on the ``framewise output'' of the output of the model, we decide whether to maintain this audio segment. For ``word segmentation'', we finetune a Cnn8-Rnn network using predefined architecture of PANNs model with pre-defined parameters. For ``subword extraction'', we change the parameters of the oscillator-based speech syllabification algorithm to make it suitable for dog sounds. For IPA vowel extraction, we collect three sets of IPA vowel sounds and use voice conversion to double the sets. Then we average the whisper feature for all 6 tracks for each ipa vowel. We calculate the Euclidien similarity between each subword and IPA vowel features to assign the most similar ipa symbol to each subword. 

In the part of ``surrounding context extraction'', when inferencing the location, we fine-tuned a Resnet50 from models pre-trained on Places365, then 5 frames are sampled uniformly from the period before and after the vocalization, and then the location category with the largest sum of logits is voted out as the final result for ``location''. As for the activity, we fine-tuned a model from TSN, then a 2-second video clip which contains 1 second before the vocalization and 1 second after the vocalization is sent to the video understanding model, and the result of the ``activity'' is inferenced based on this clip. 

The whole experiment is running on an Ubuntu system server with four 2080ti GPUs and 128G RAM. 

\section{Location and Activity Inference Error Analysis}

In this part, we will evaluate the methods we raised which can tell us where the dog is (location) and what the dog is doing (activity). 

\begin{figure*}[htb]
	\centering
	\begin{subfigure}[]{0.4\textwidth}
		\centering
		\includegraphics[width=0.9\textwidth]{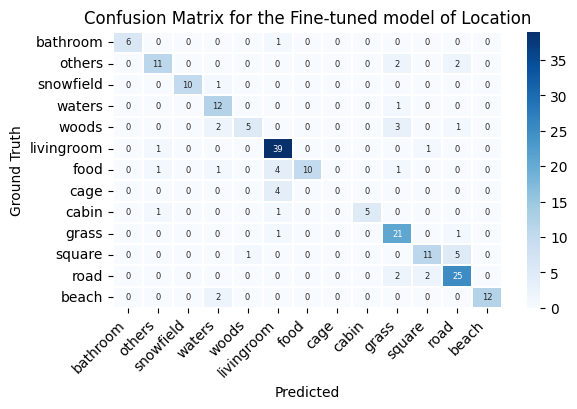}
		\caption{Location model confusion matrix.}
		\label{fig:context evaluation 1}
	\end{subfigure}
	\begin{subfigure}[]{0.4\textwidth}
		\centering
		\includegraphics[width=0.9\textwidth]{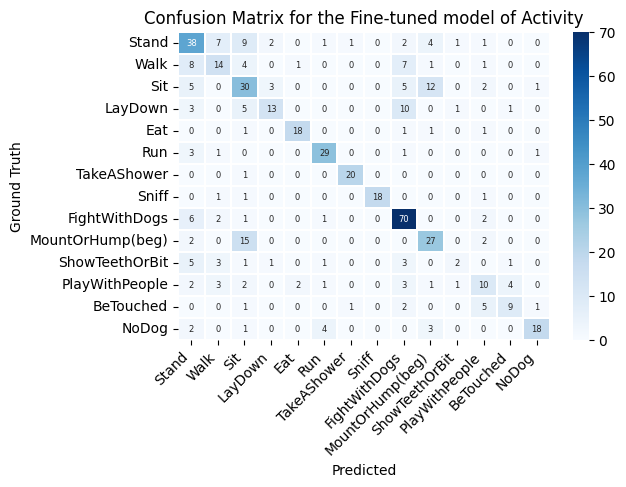}
		\caption{Activity model confusion matrix.}
		\label{fig:context evaluation 2}
	\end{subfigure}
\caption{Context Evaluation.}
\label{fig:context evaluation}
\end{figure*}

To get the location from the video, we fine-tuned an image classification model and vote on 5 frame results to decide the location. 
The overall accuracy of this model is 77.99\%. ~\figref{fig:context evaluation 1} presents the confusion matrix of a test set result with 209 images, the unit of the number in each cell is one image sample. For most of the location classes, this model can precisely infer the location. There may be some confusion between categories, such as ``food'' and ``living room'', ``water'' and ``woods'', this may be due to the simultaneous appearance of the characteristics of multiple locations in the same picture. For the class ``cage'', the recall is a bit low from this confusion matrix. One possible reason is that there are only 4 samples from this category. From the further exploration of the dataset, the model can still distinguish ``cage'' from ``living room''. 

To get the activity from the video, we fine-tuned a video understanding model, and use the model to decide the dog activity based on a 2-seconds period of video. The overall accuracy of this model is 61.40\%. ~\figref{fig:context evaluation 2} presents the confusion matrix of a test set result with 513 videos on this model. There are some categories that can be confused with each other, like ``sit'' and ``mount or hump (beg)'', ``stand'' and ``walk'', ``play with people'' and ``be touched''. The reason that these pairs can lead to more confusion is that the activities themself are similar, this realization is reflected in the different performance of the model across categories.

\section{Details of Dataset}

\begin{table}[th]
\small
\centering
\begin{tabular}{p{0.3\columnwidth}|p{0.3\columnwidth}}
\toprule
\textbf{Word Type} & \textbf{Duration} \\
\hline
Whimper & 3.32 seconds \\
\hline
Yip & 0.36 seconds \\
\hline
Bow-wow & 0.28 seconds\\
\hline
Growl & 0.88 seconds \\
\hline
Bark & 0.4 seconds \\
\hline
Howl & 1.08 seconds \\
\bottomrule
\end{tabular}
\caption{Summary of duration for word types.}
\label{tab:duration}
\end{table}

We present the average duration and stand deviation of each ipa vowel symbol assigned to subwords in ~\tabref{tab:ipa_duration}. The duration of each ipa symbol in different word types varies from each other which is illustrated by time standard deviation. 
\begin{figure}[th]
\centering
\includegraphics[width=0.8\columnwidth]{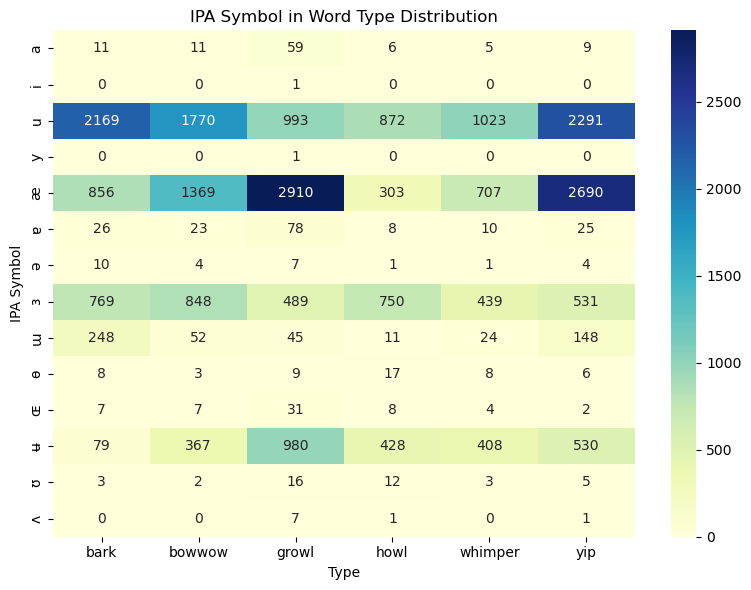}
\caption{IPA symbol in word type distribution.} 
\label{fig:ipa_type_distri}
\end{figure}

\begin{table}[th]
\small
\centering
\begin{tabular}{p{0.2\columnwidth}|p{0.3\columnwidth}|p{0.4\columnwidth}}
\toprule
\textbf{IPA Vowel Symbol} & \textbf{Duration} & \textbf{Standard Deviation of Time Duration for Word Types} \\
\hline
u & 0.293 seconds & 0.0276 \\
\hline
\textrevepsilon & 0.213 seconds & 0.0312 \\
\hline
æ & 0.211 seconds & 0.0171\\
\hline
\textbaru & 0.233 seconds & 0.0152\\
\hline
\textturnm & 0.216 seconds & 0.0492\\
\hline
\textscoelig & 0.422 seconds & 0.0293\\
\hline
\textsci & 0.311 seconds & 0.0216\\
\hline
\textbaro & 0.479 seconds &0.0830\\
\hline
\textturna & 0.409 seconds &0.0217\\
\hline
a & 0.516 seconds & 0.0318 \\
\hline
\textschwa & 0.582 seconds & 0.0606 \\
\hline
\textturnv & 0.555 seconds & 0.0209\\
\hline
y & 0.528 seconds & 0 (only occurs one time)\\
\hline
i & 0.335 seconds & 0 (only occurs one time)\\
\bottomrule
\end{tabular}
\caption{Summary of duration and standard deviation for ipa vowel symbols.}
\label{tab:ipa_duration}
\end{table}

\section{Experiments for Analysing Subwords}

\begin{figure*}[h]
\centering
\begin{subfigure}[]{0.4\textwidth}
	\centering
	\includegraphics[width=0.9\textwidth]{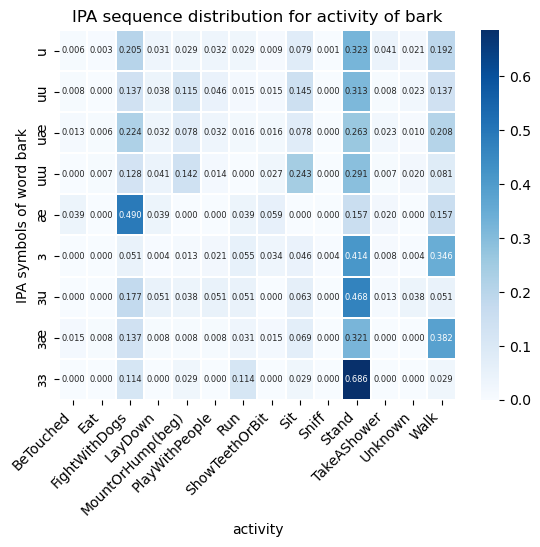}
	\caption{Activity distribution of subwords for bark.}
	\label{fig:1}
\end{subfigure}
\begin{subfigure}[]{0.4\textwidth}
	\centering
	\includegraphics[width=0.9\textwidth]{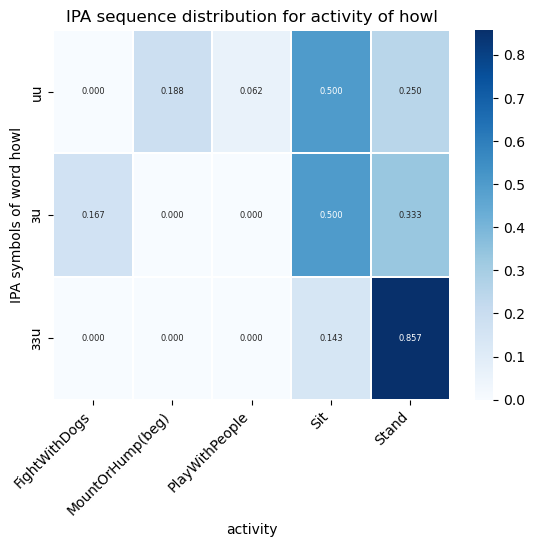}
	\caption{Activity distribution of subwords for howl.}
	\label{fig:2}
\end{subfigure}

\begin{subfigure}[]{0.4\textwidth}
	\centering
	\includegraphics[width=0.9\textwidth]{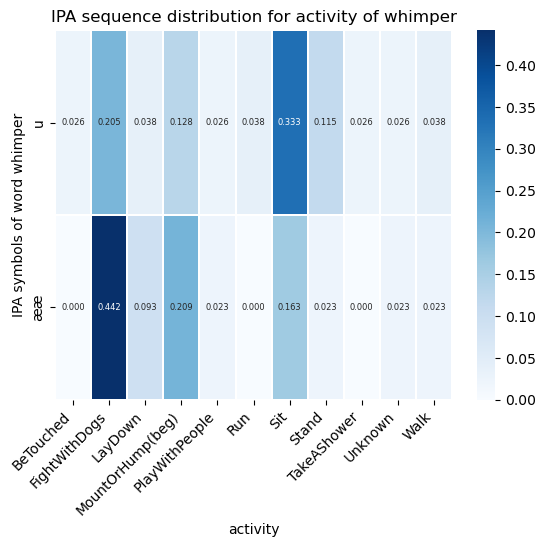}
	\caption{Activity distribution of subwords for whimper.}
\label{fig:3}
\end{subfigure}
\begin{subfigure}[]{0.4\textwidth}
	\centering
	\includegraphics[width=0.9\textwidth]{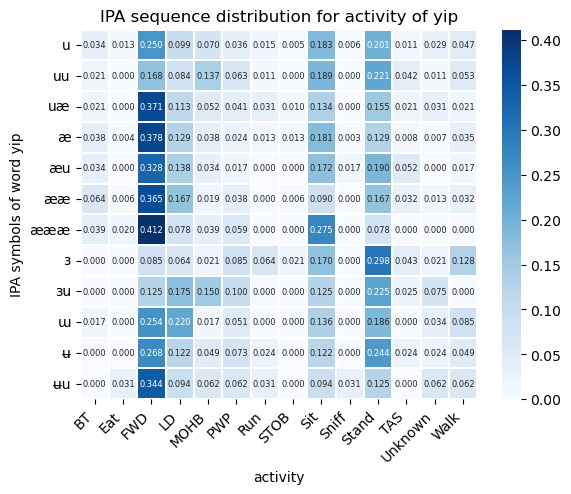}
	\caption{Activity distribution of subwords for yip.}
\label{fig:4}
\end{subfigure}

\begin{subfigure}[]{0.4\textwidth}
	\centering
	\includegraphics[width=0.9\textwidth]{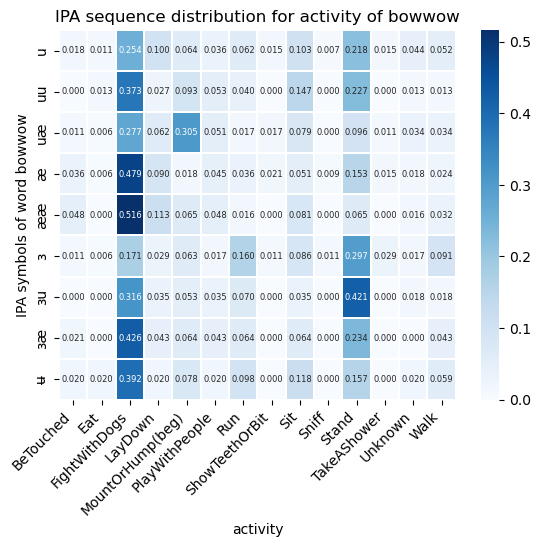}
	\caption{Activity distribution of subwords for bowwow.}
\label{fig:5}
\end{subfigure}
\begin{subfigure}[]{0.4\textwidth}
	\centering
	\includegraphics[width=0.9\textwidth]{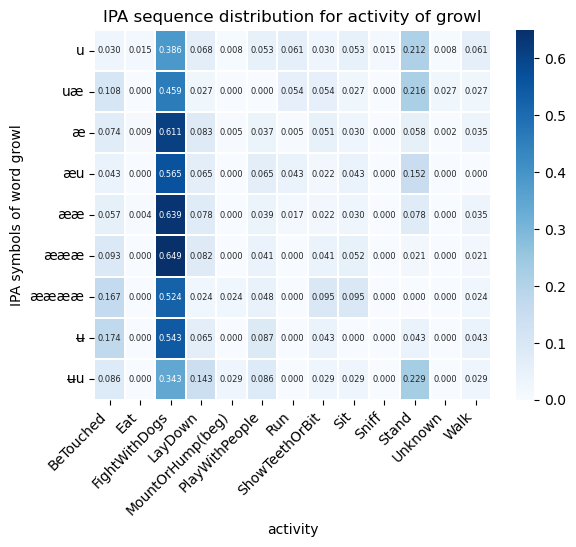}
	\caption{Activity distribution of subwords for growl.}
	\label{fig:6}
\end{subfigure}

\caption{Activity distribution of subwords.}
\label{fig:activity_subwords}
\end{figure*}

\begin{table}[t!]
\small
\centering
\begin{tabular}{p{0.45\columnwidth}|p{0.45\columnwidth}}
\toprule
\textbf{Word Type for Top 5 Frequent Subwords} & \textbf{Number of Dog Used} \\
\hline
Yip & 17.6 \\
\hline
Whimper & 9 \\
\hline
Howl & 4.25\\
\hline
Growl & 17 \\
\hline
Bowwow & 24.6 \\
\hline
Bark & 19.6 \\
\bottomrule
\end{tabular}
\caption{Summary to show frequent subwords are used by multiple dogs.}
\label{tab:subwords}
\end{table}

We collect the average number of different dogs using this subword for the 5 most frequent subwords of each word type in ~\tabref{tab:subwords}. ~\figref{fig:ipa_type_distri} shows the number of IPA symbols in different word types.
In order to prove that a word type can be further divided into more fine-grained types, for each word type, we collect the activity distribution of its IPA transcript which is a sequence of ipa symbols assigned to subwords. We find that the data distribution of word type ipa transcript can be mainly divided into several types, as they show an incoherent distribution for activities which implies in the same word type, these subparts convey different semantics. This shows that these words can be further divided into several different types. ~\figref{fig:activity_subwords} present the activity distribution for frequent ipa transcript for each word type as it is more intuitive using activity. 

\begin{figure*}[h]
\centering
\begin{subfigure}[]{0.4\textwidth}
	\centering
	\includegraphics[width=0.9\textwidth]{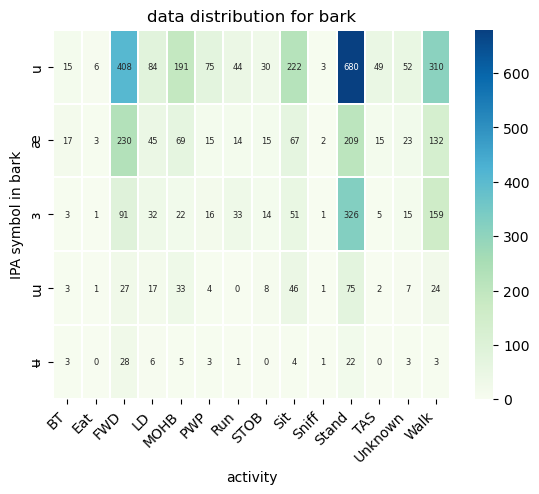}
	\caption{Activity distribution of subwords for bark.}
	\label{fig:11}
\end{subfigure}
\begin{subfigure}[]{0.4\textwidth}
	\centering
	\includegraphics[width=0.9\textwidth]{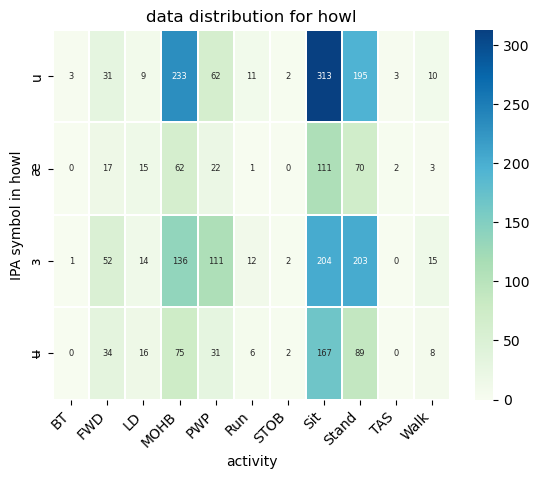}
	\caption{Activity distribution of subwords for howl.}
	\label{fig:21}
\end{subfigure}

\begin{subfigure}[]{0.4\textwidth}
	\centering
	\includegraphics[width=0.9\textwidth]{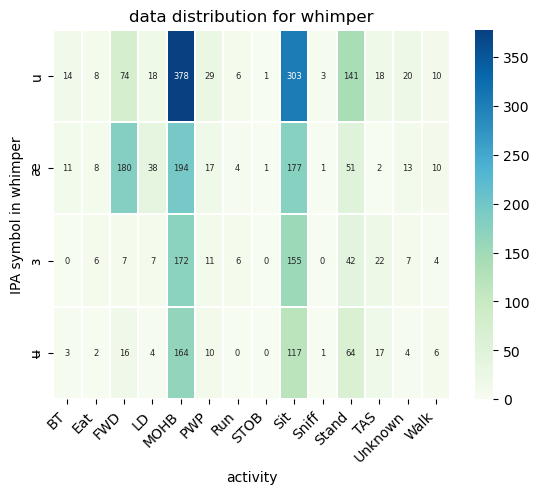}
	\caption{Activity distribution of subwords for whimper.}
\label{fig:31}
\end{subfigure}
\begin{subfigure}[]{0.4\textwidth}
	\centering
	\includegraphics[width=0.9\textwidth]{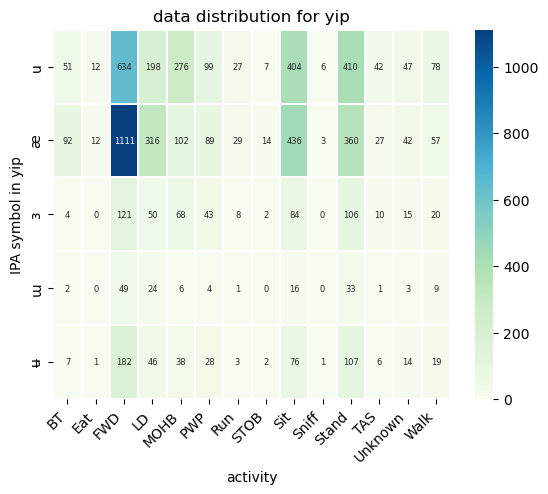}
	\caption{Activity distribution of subwords for yip.}
\label{fig:41}
\end{subfigure}

\begin{subfigure}[]{0.4\textwidth}
	\centering
	\includegraphics[width=0.9\textwidth]{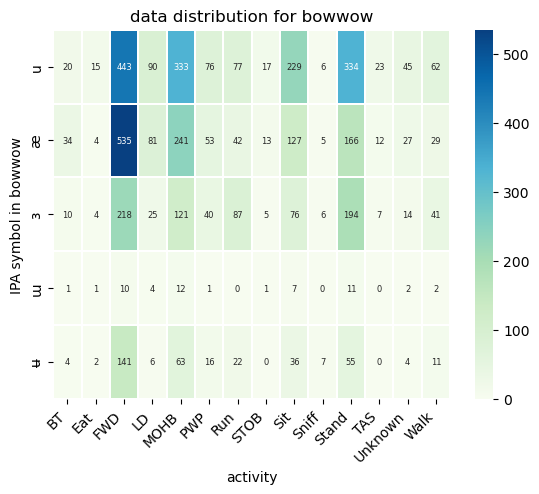}
	\caption{Activity distribution of subwords for bowwow.}
\label{fig:51}
\end{subfigure}
\begin{subfigure}[]{0.4\textwidth}
	\centering
	\includegraphics[width=0.9\textwidth]{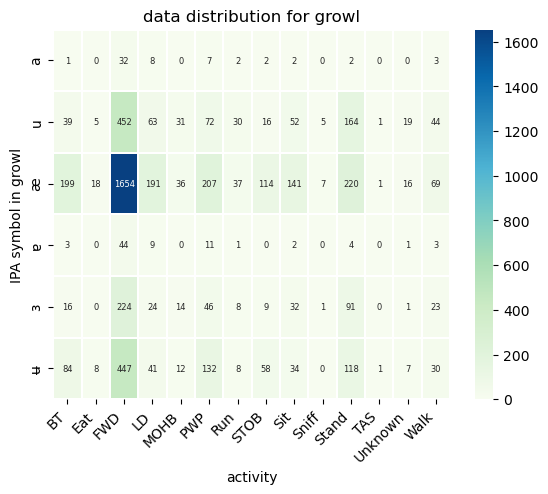}
	\caption{Activity distribution of subwords for growl.}
	\label{fig:61}
\end{subfigure}

\caption{Activity distribution of IPA symbol in word type.}
\label{fig:activity_symbol_word_type}
\end{figure*}

If the subword conveys semantics, we can observe that these IPA symbols should present a different context distribution compared with word distribution.   For further analysis, we present IPA symbol distribution of activity for each word type in ~\figref{fig:activity_symbol_word_type}. We use activity instead of context and simply plot the number of coexistence instead of normalizing for clearness. 

\section{Example scenario for acitivity and location}
\begin{figure*}[h]
\centering
\begin{subfigure}[]{0.4\textwidth}
	\centering
	\includegraphics[width=0.9\textwidth]{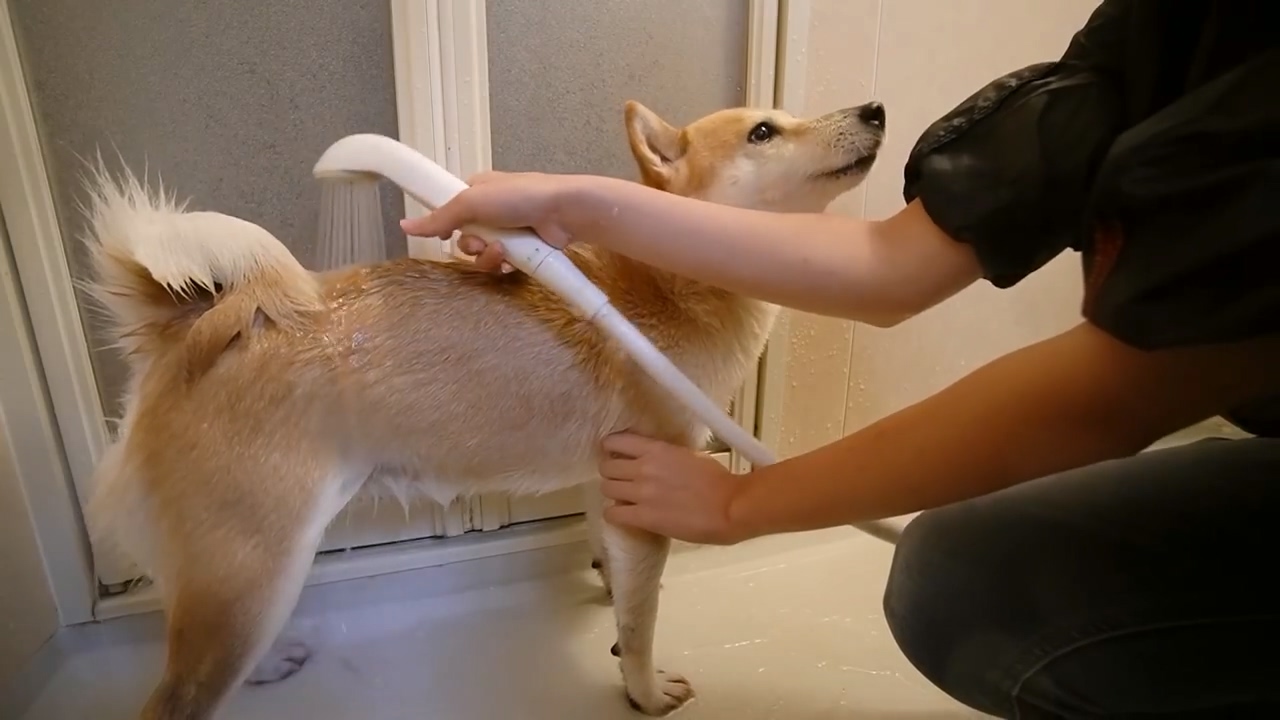}
	\caption{Bathroom example.}
	\label{fig:loc1}
\end{subfigure}
\begin{subfigure}[]{0.4\textwidth}
	\centering
	\includegraphics[width=0.9\textwidth]{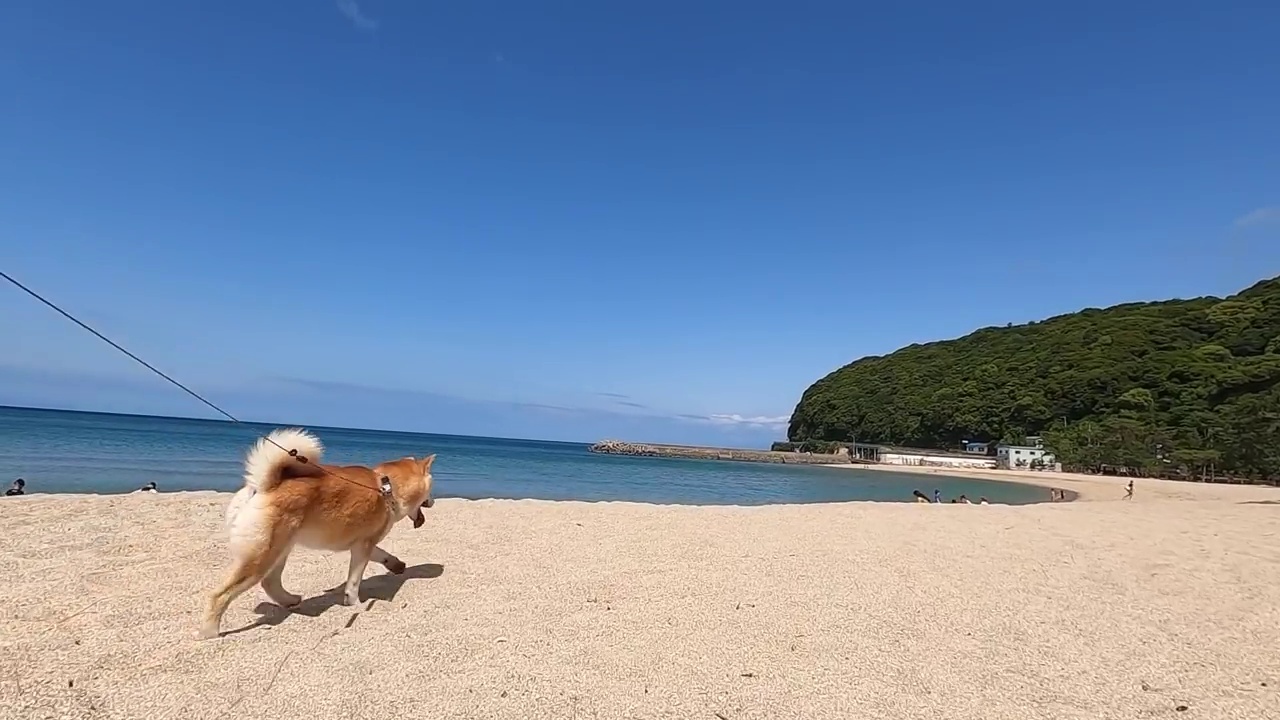}
	\caption{Beach example.}
	\label{fig:loc2}
\end{subfigure}

\begin{subfigure}[]{0.4\textwidth}
	\centering
	\includegraphics[width=0.9\textwidth]{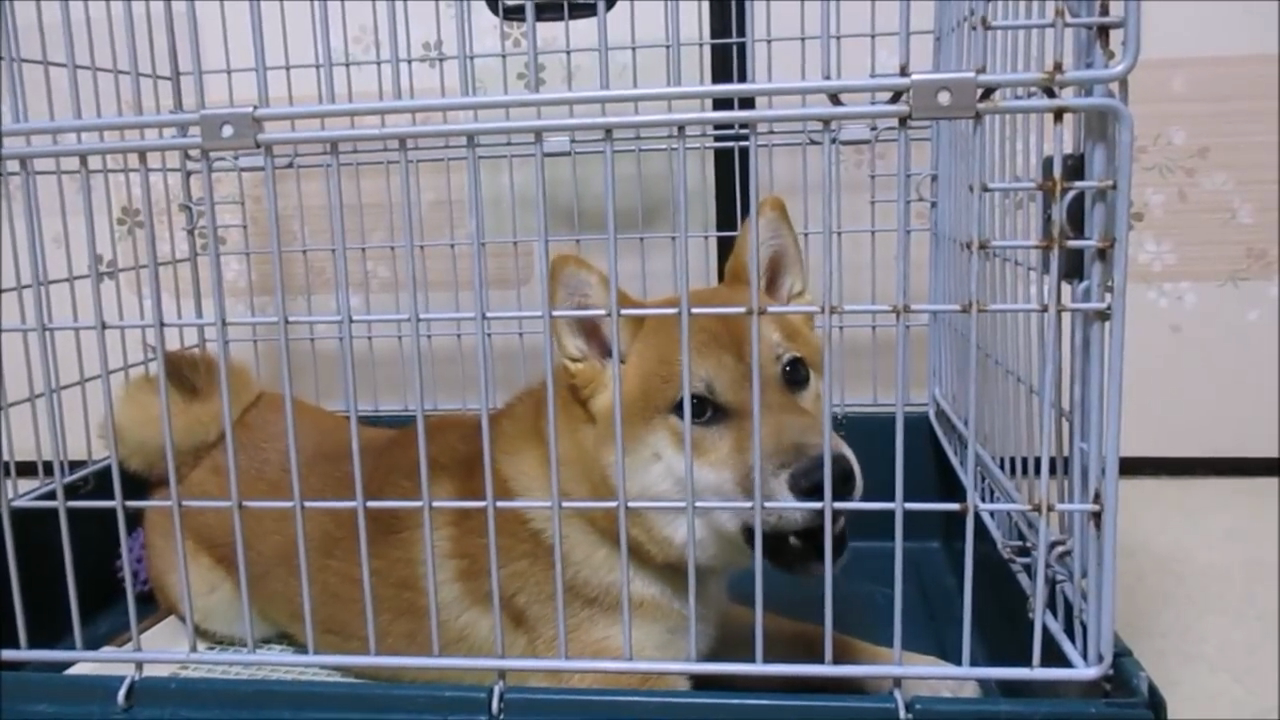}
	\caption{Cage example.}
\label{fig:loc3}
\end{subfigure}
\begin{subfigure}[]{0.4\textwidth}
	\centering
	\includegraphics[width=0.9\textwidth]{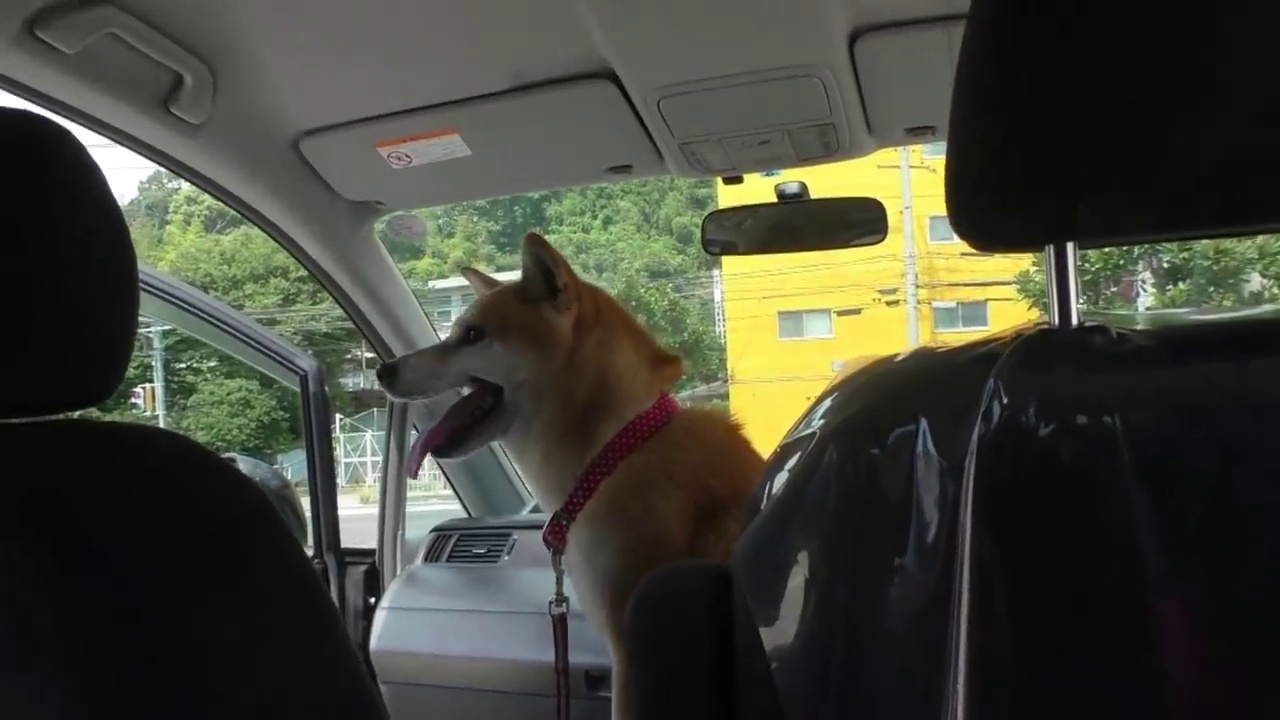}
	\caption{Vehical cabin example.}
\label{fig:loc4}
\end{subfigure}

\begin{subfigure}[]{0.4\textwidth}
	\centering
	\includegraphics[width=0.9\textwidth]{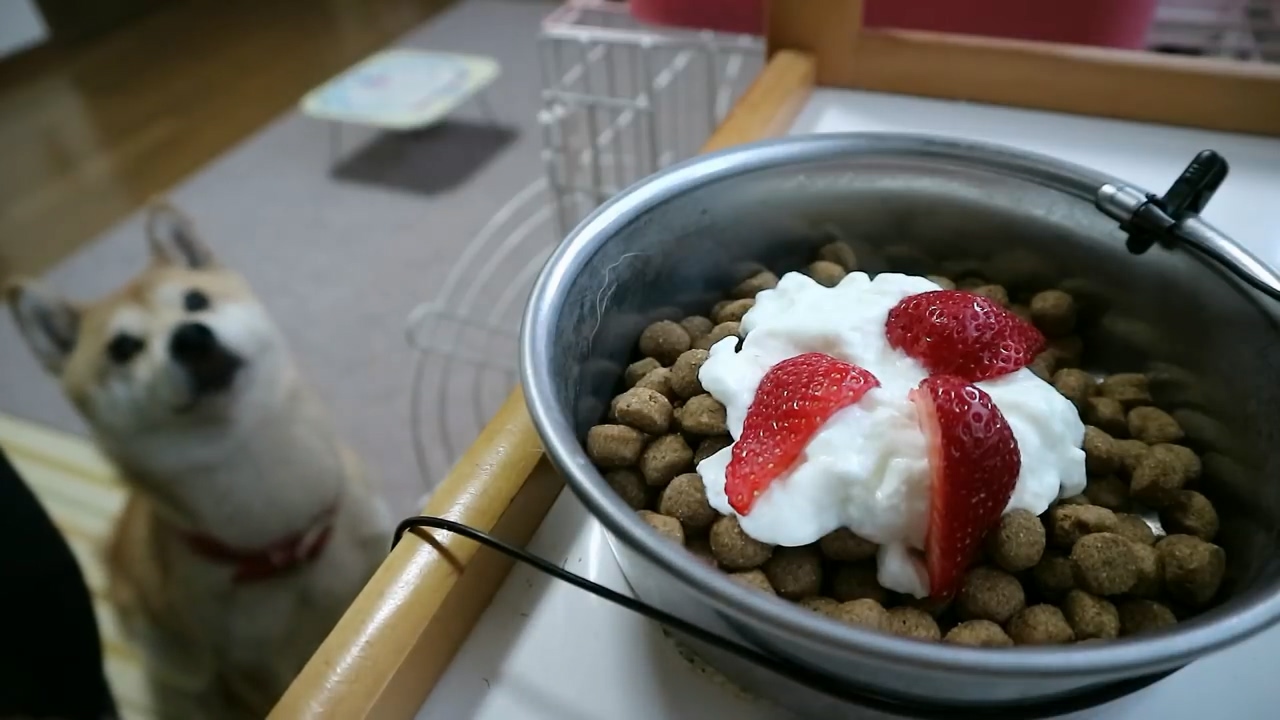}
	\caption{Food nearby example.}
\label{fig:loc5}
\end{subfigure}
\begin{subfigure}[]{0.4\textwidth}
	\centering
	\includegraphics[width=0.9\textwidth]{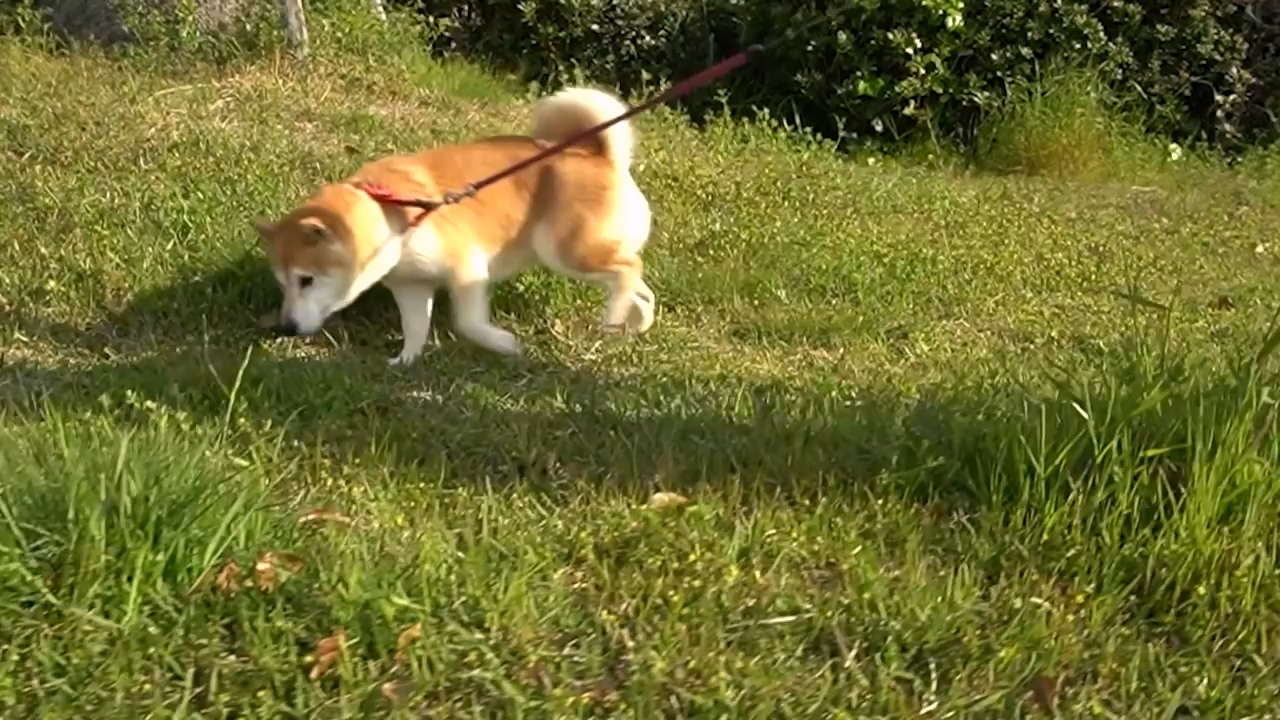}
	\caption{Grass example.}
	\label{fig:loc6}
\end{subfigure}

\begin{subfigure}[]{0.4\textwidth}
	\centering
	\includegraphics[width=0.9\textwidth]{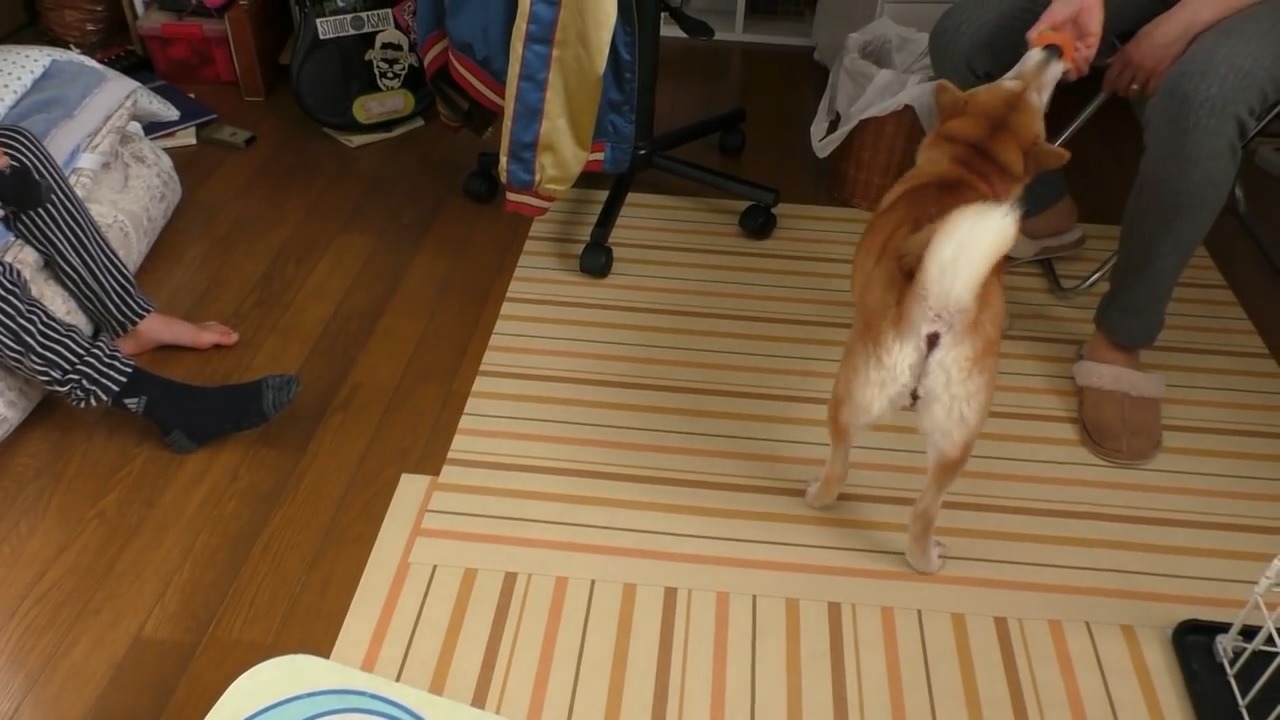}
	\caption{Living room example.}
\label{fig:loc7}
\end{subfigure}
\begin{subfigure}[]{0.4\textwidth}
	\centering
	\includegraphics[width=0.9\textwidth]{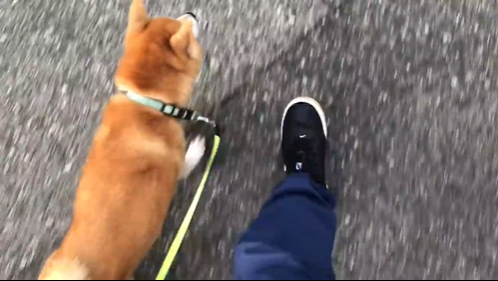}
	\caption{Road example.}
	\label{fig:loc8}
\end{subfigure}

\begin{subfigure}[]{0.4\textwidth}
	\centering
	\includegraphics[width=0.9\textwidth]{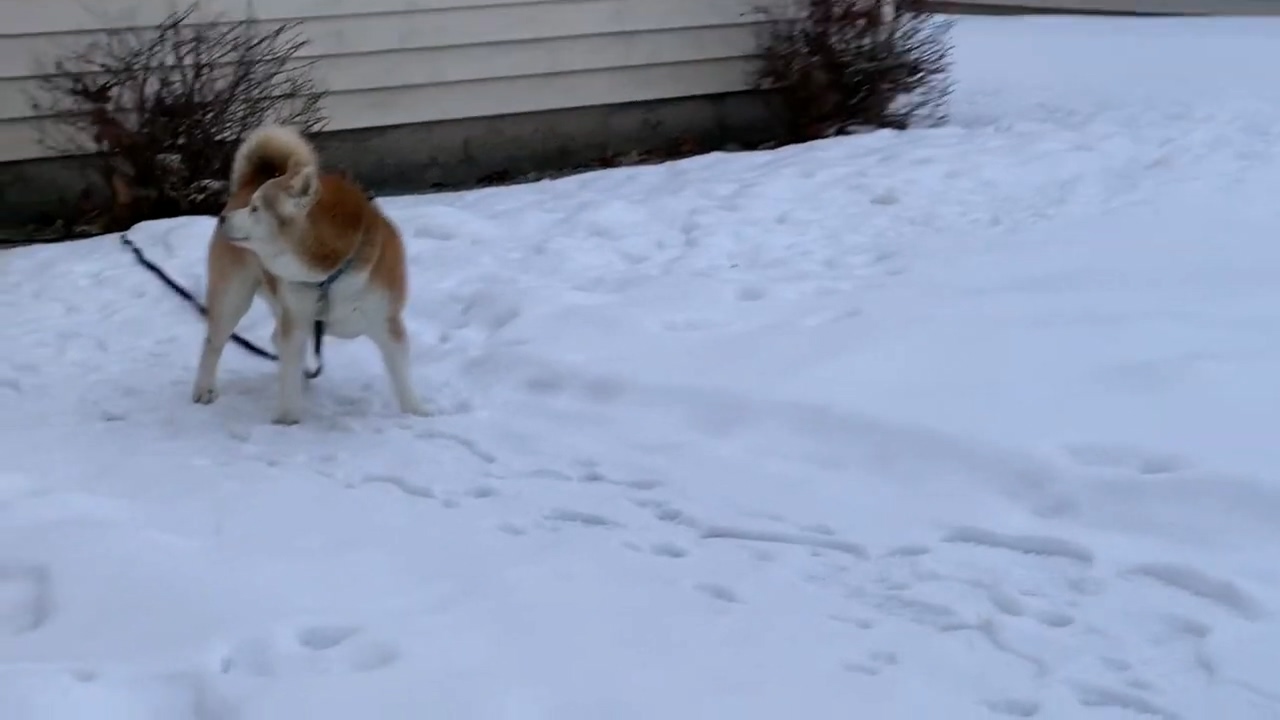}
	\caption{Snowfield example.}
\label{fig:loc9}
\end{subfigure}
\begin{subfigure}[]{0.4\textwidth}
	\centering
	\includegraphics[width=0.9\textwidth]{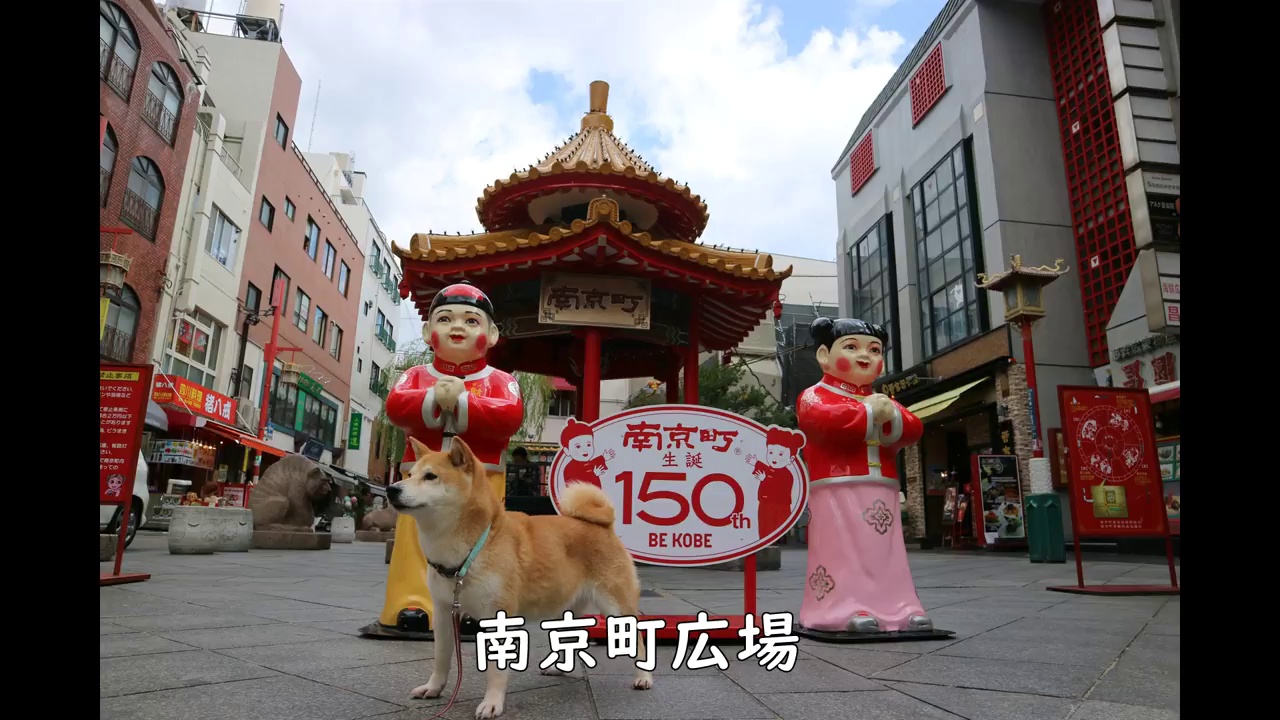}
	\caption{Square example.}
	\label{fig:loc10}
\end{subfigure}

\begin{subfigure}[]{0.4\textwidth}
	\centering
	\includegraphics[width=0.9\textwidth]{images/snowfield.jpg}
	\caption{Snowfield example.}
\label{fig:loc11}
\end{subfigure}

\caption{Example for locations.}
\label{fig:location_example}
\end{figure*}

\begin{figure*}[h]
\centering
\begin{subfigure}[]{0.3\textwidth}
	\centering
	\includegraphics[width=0.9\textwidth]{images/mountorhump(beg).jpg}
	\caption{Mount or hump (beg) example.}
	\label{fig:act1}
\end{subfigure}
\begin{subfigure}[]{0.3\textwidth}
	\centering
	\includegraphics[width=0.9\textwidth]{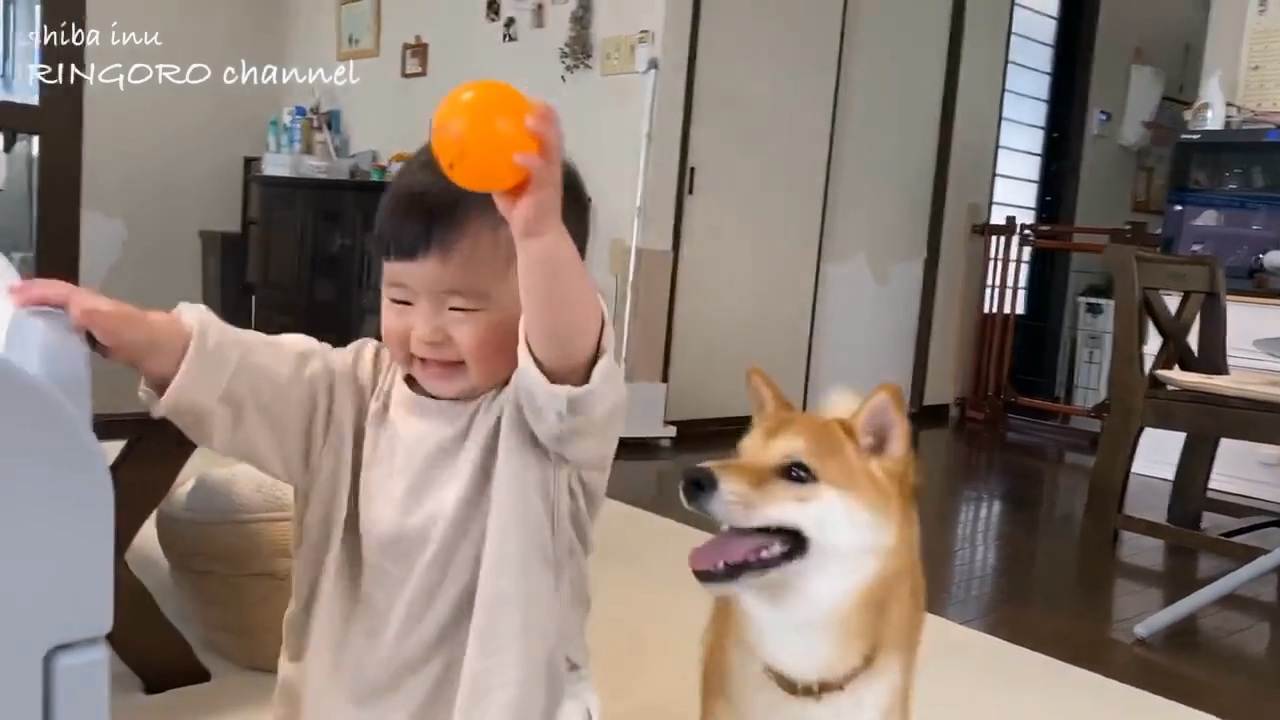}
	\caption{Play with people example.}
	\label{fig:act2}
\end{subfigure}
\begin{subfigure}[]{0.3\textwidth}
	\centering
	\includegraphics[width=0.9\textwidth]{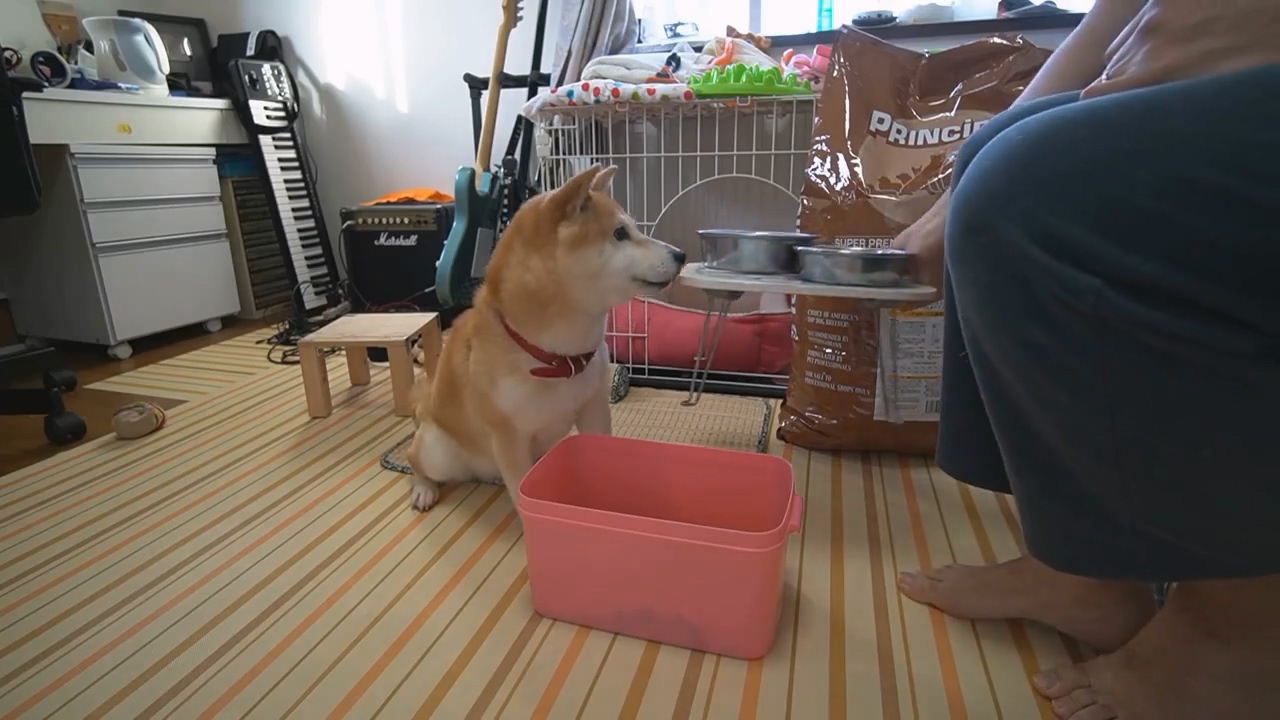}
	\caption{Sit example.}
\label{fig:act3}
\end{subfigure}

\begin{subfigure}[]{0.3\textwidth}
	\centering
	\includegraphics[width=0.9\textwidth]{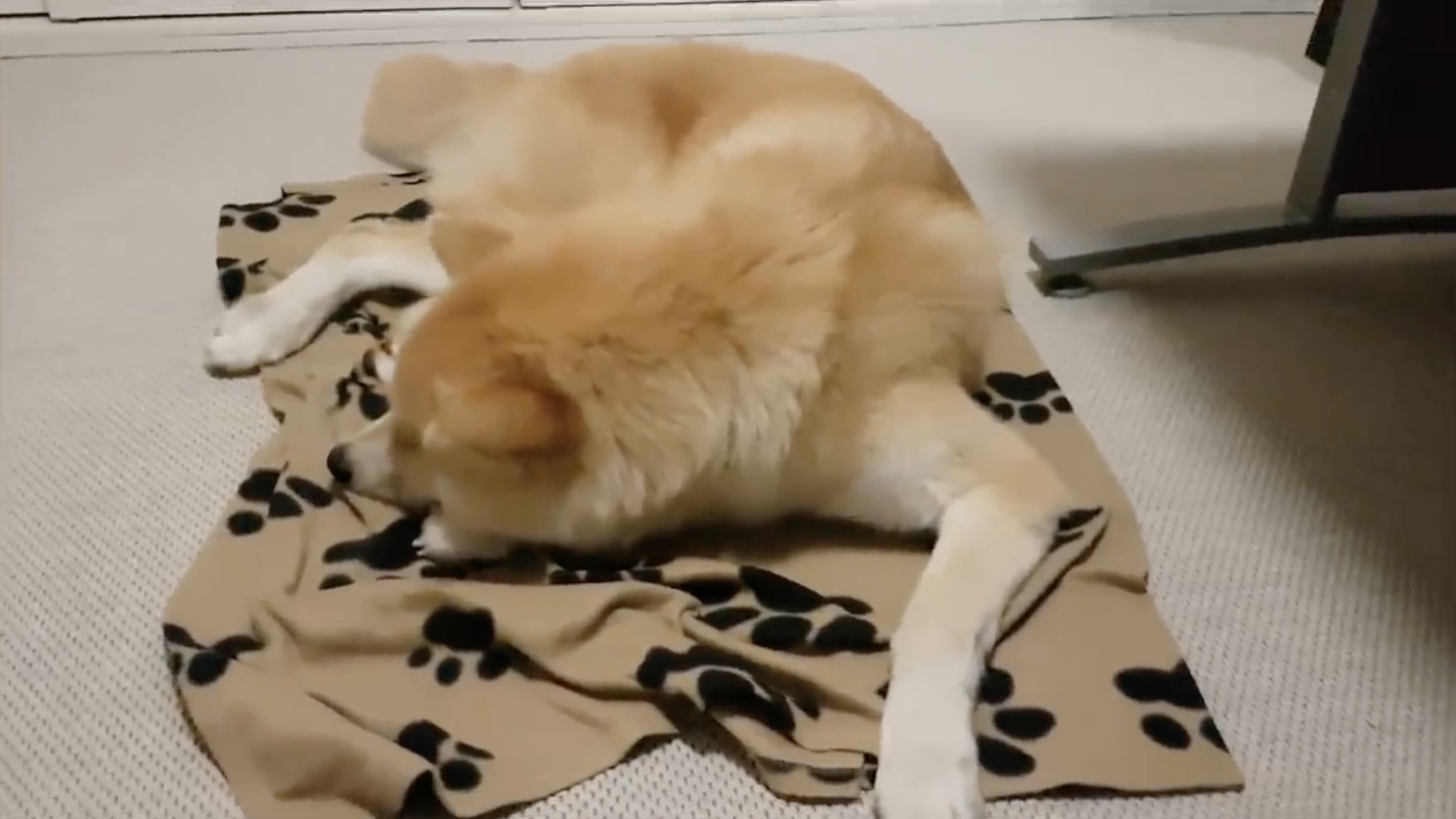}
	\caption{Laydown example.}
\label{fig:act4}
\end{subfigure}
\begin{subfigure}[]{0.3\textwidth}
	\centering
	\includegraphics[width=0.9\textwidth]{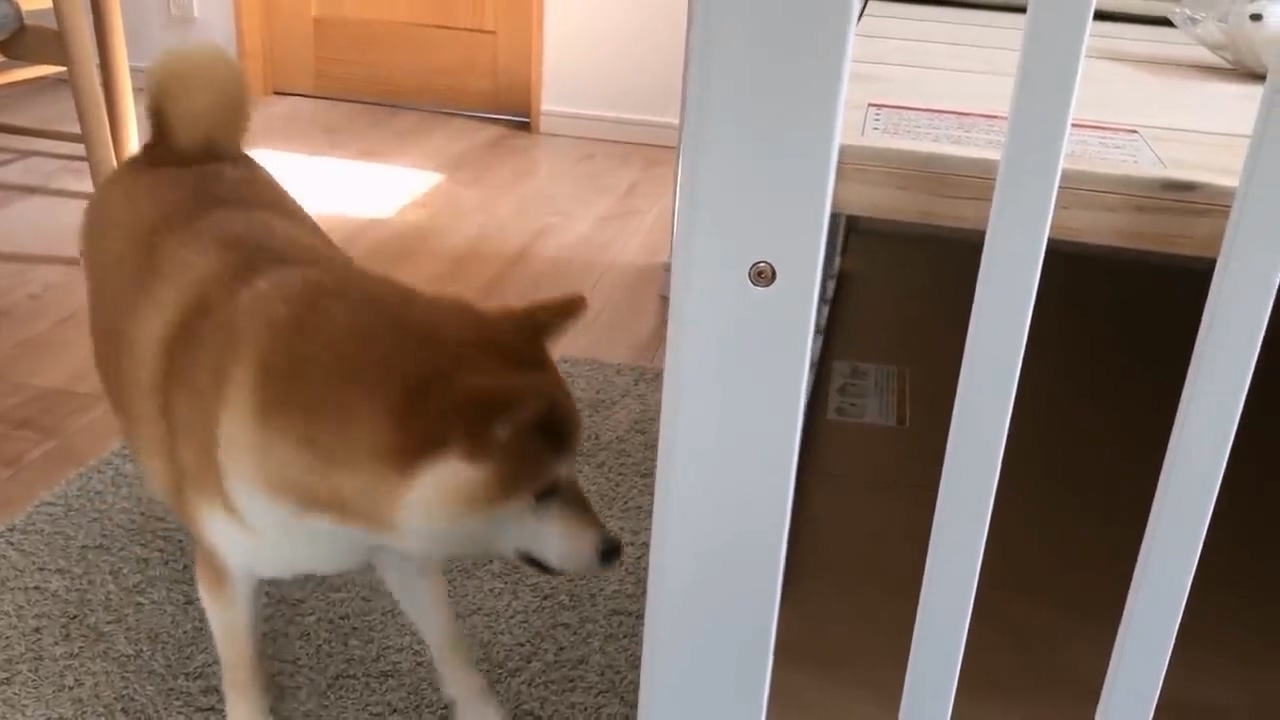}
	\caption{Walk example.}
\label{fig:act5}
\end{subfigure}
\begin{subfigure}[]{0.3\textwidth}
	\centering
	\includegraphics[width=0.9\textwidth]{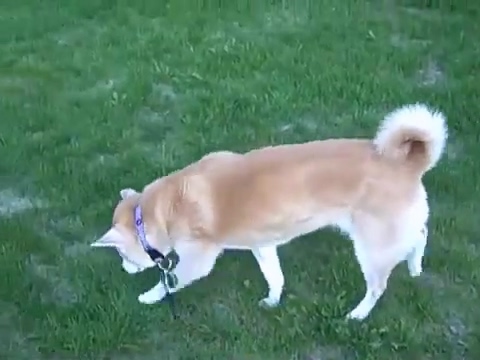}
	\caption{Sniff example.}
	\label{fig:act6}
\end{subfigure}

\begin{subfigure}[]{0.3\textwidth}
	\centering
	\includegraphics[width=0.9\textwidth]{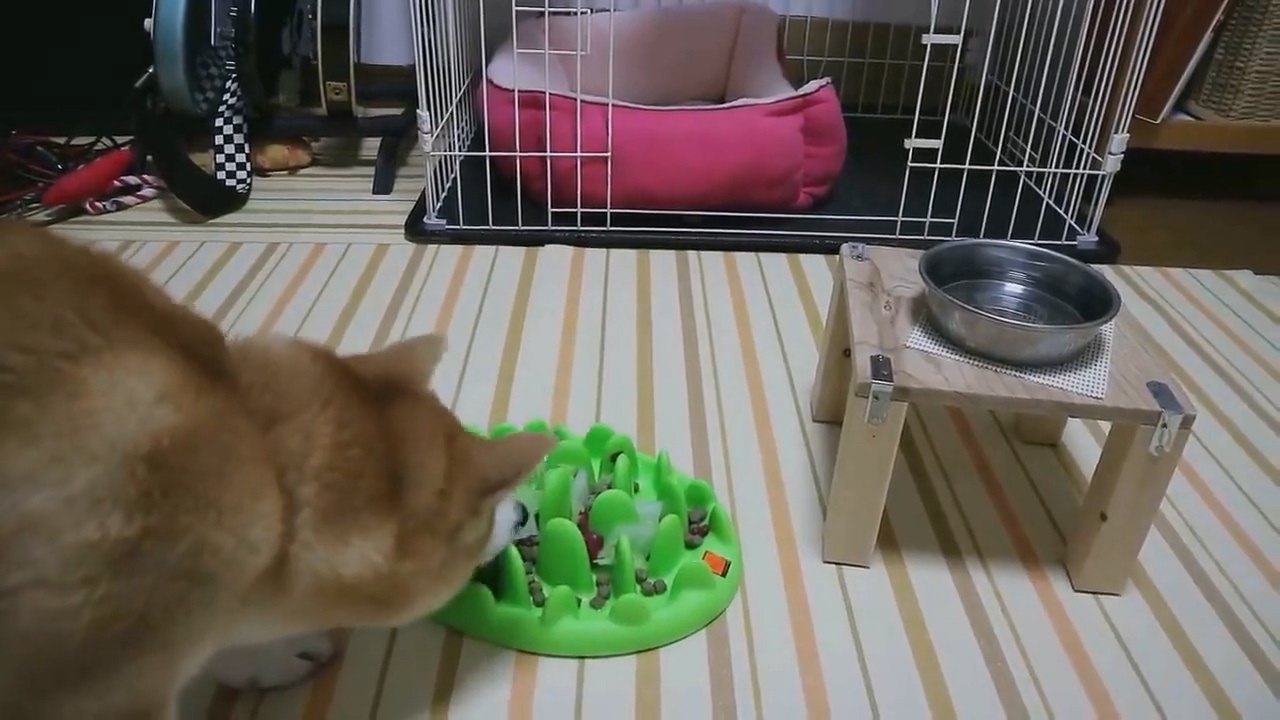}
	\caption{Eat example.}
\label{fig:act7}
\end{subfigure}
\begin{subfigure}[]{0.3\textwidth}
	\centering
	\includegraphics[width=0.9\textwidth]{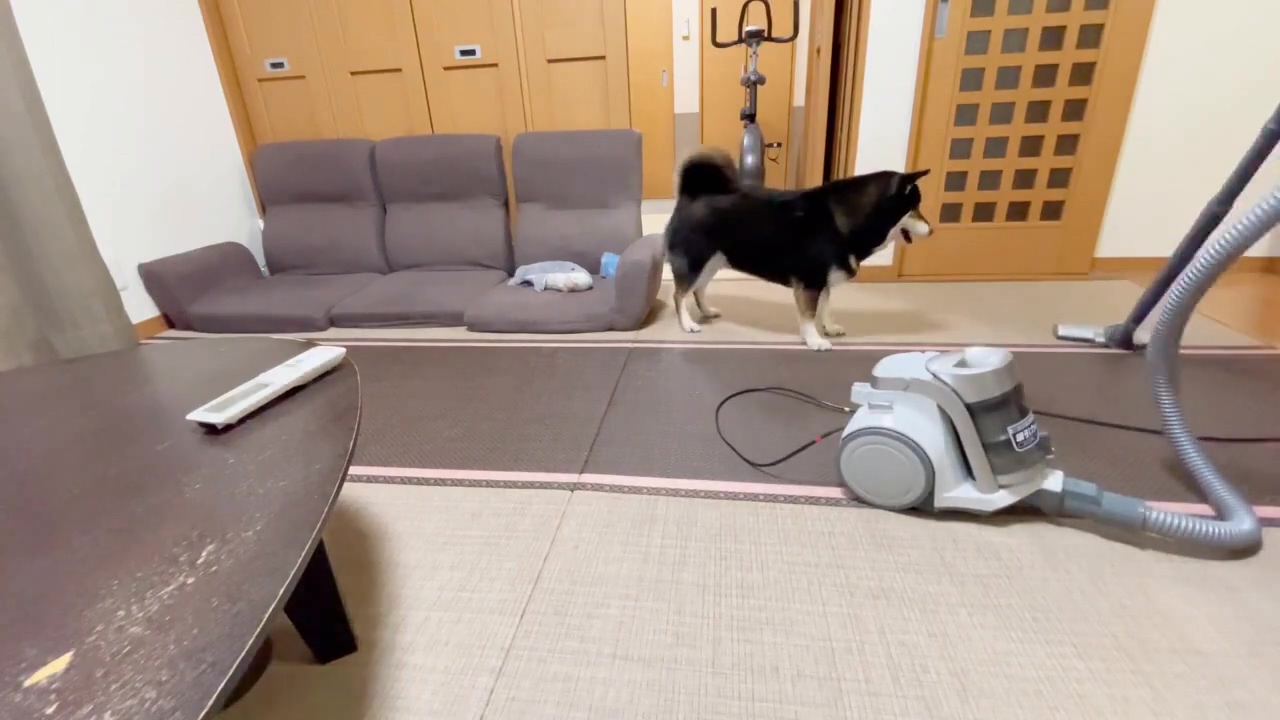}
	\caption{Stand example.}
	\label{fig:act8}
\end{subfigure}
\begin{subfigure}[]{0.3\textwidth}
	\centering
	\includegraphics[width=0.9\textwidth]{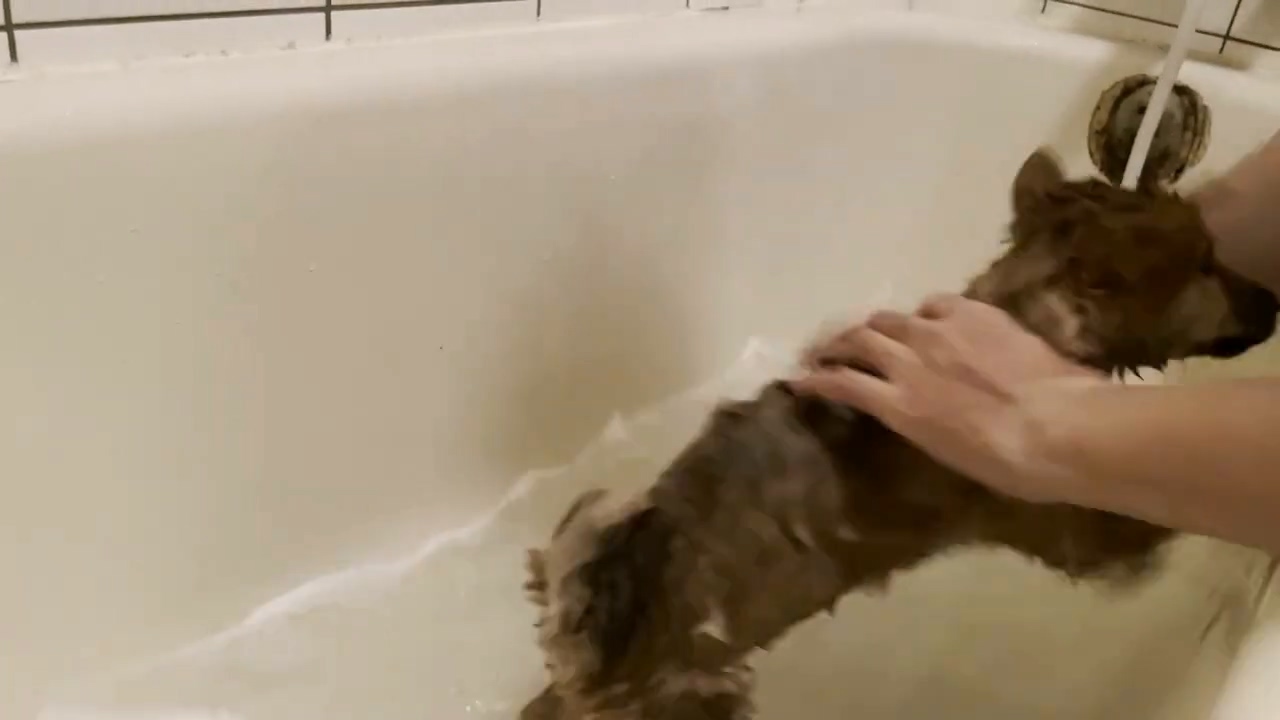}
	\caption{Take a shower example.}
\label{fig:act9}
\end{subfigure}

\begin{subfigure}[]{0.3\textwidth}
	\centering
	\includegraphics[width=0.9\textwidth]{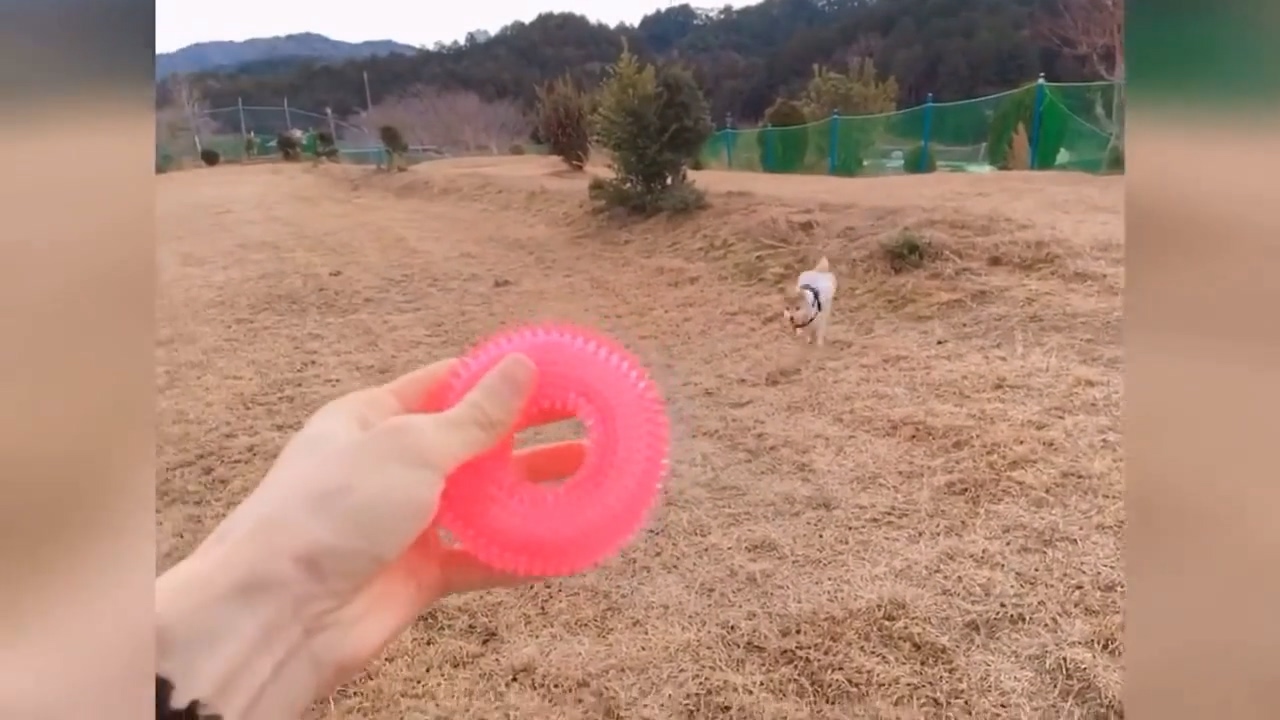}
	\caption{Run example.}
	\label{fig:act10}
\end{subfigure}
\begin{subfigure}[]{0.3\textwidth}
	\centering
	\includegraphics[width=0.9\textwidth]{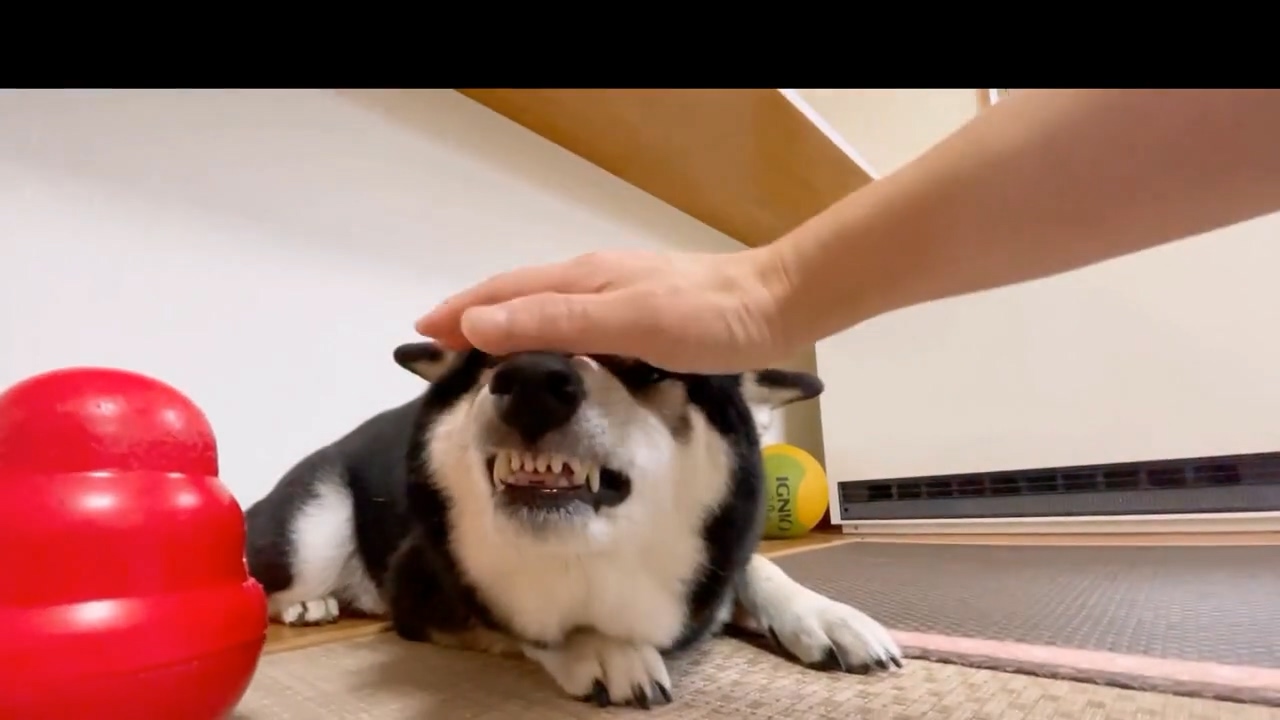}
	\caption{Be touched example.}
\label{fig:act11}
\end{subfigure}
\begin{subfigure}[]{0.3\textwidth}
	\centering
	\includegraphics[width=0.9\textwidth]{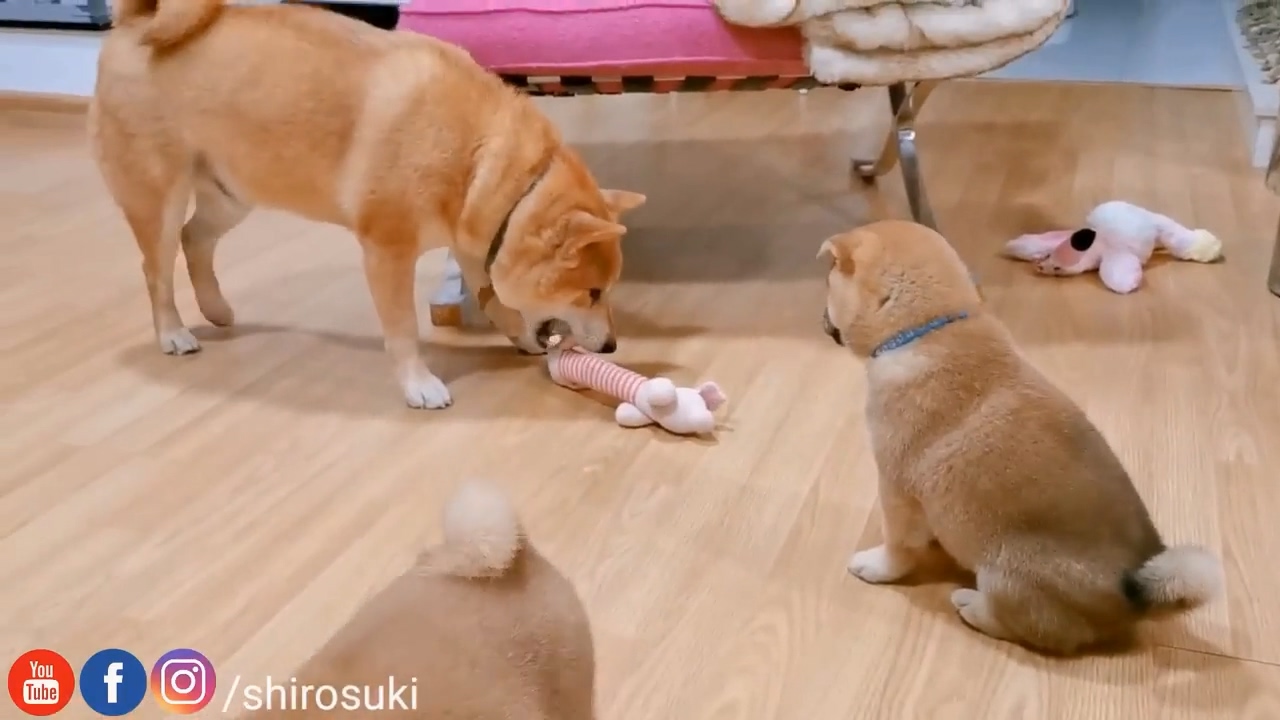}
	\caption{Show teech or bit example.}
\label{fig:act12}
\end{subfigure}

\begin{subfigure}[]{0.3\textwidth}
	\centering
	\includegraphics[width=0.9\textwidth]{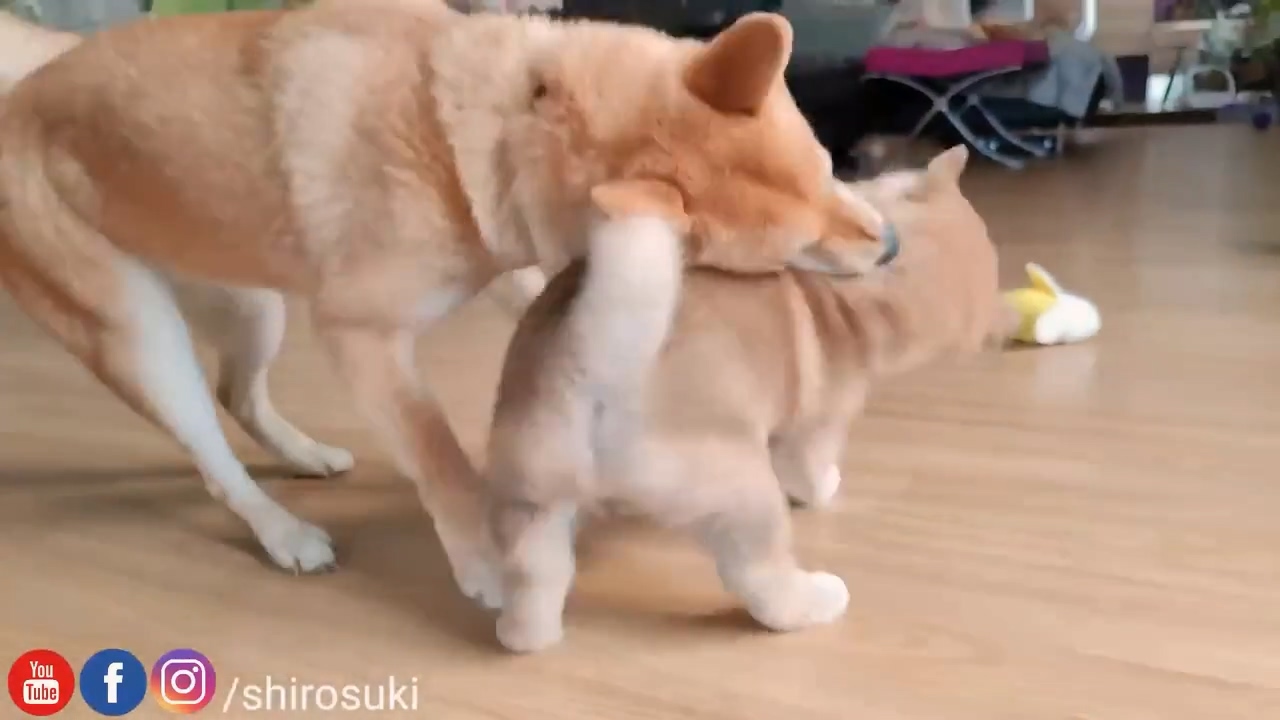}
	\caption{Fight with dogs example.}
	\label{fig:act13}
\end{subfigure}

\caption{Example for activities.}
\label{fig:activity_example}
\end{figure*}

In this section, we present picture examples to illustrate each activity and location for better understanding. ~\figref{fig:location_example} present the examples for locations. ~\figref{fig:activity_example} present the examples for activities. 

